\documentclass{article}

\usepackage{PRIMEarxiv}

\usepackage[utf8]{inputenc} 
\usepackage[T1]{fontenc}    
\usepackage{hyperref}       
\usepackage{url}            
\usepackage{booktabs}       
\usepackage{amsfonts}       
\usepackage{nicefrac}       
\usepackage{microtype}      
\usepackage{lipsum}
\usepackage{fancyhdr}       
\usepackage{graphicx}       
\graphicspath{{media/}}     

\usepackage{multirow}
\usepackage{array}
\newcolumntype{P}[1]{>{\centering\arraybackslash}p{#1}}
\newcolumntype{M}[1]{>{\centering\arraybackslash}m{#1}}
\usepackage{graphicx}
\usepackage{amsmath}
\usepackage{amssymb}
\usepackage{float}
\usepackage{tcolorbox}
\usepackage{xcolor}

\title{A Multi-Model Probabilistic Framework for Seismic Risk Assessment and Retrofit Planning of Electric Power Networks
}

\author{
  Huangbin Liang\\
  Future Resilient Systems \\Singapore-ETH Centre \\
  Singapore\\
Corresponding email:\\ huangbin.liang@sec.ethz.ch\\
   \And
  Beatriz Moya \\
  ENSAM Institute of Technology \\
  Paris, France\\
  \\
  CNRS@CREATE LTD., CNRS \\
  Singapore\\
  \AND
   Francisco Chinesta\\
  ENSAM Institute of Technology \\
  Paris, France\\
    \\
  CNRS@CREATE LTD., CNRS \\
  Singapore\\
  \And
  Eleni Chatzi \\
  Department of Civil Environmental \\ and Geomatic Engineering \\
  ETH Zürich \\
  Zurich, Switzerland\\
}

\begin{document}
\maketitle

\begin{abstract}
Electric power networks are critical lifelines, and their disruption during earthquakes can lead to severe cascading failures and significantly hinder post-disaster recovery. Enhancing their seismic resilience requires identifying and strengthening vulnerable components in a cost-effective and system-aware manner. However, existing studies often overlook the systemic behavior of power networks under seismic loading. Common limitations include isolated component analyses that neglect network-wide interdependencies, oversimplified damage models assuming binary states or damage independence, and the exclusion of electrical operational constraints. These simplifications can result in inaccurate risk estimates and inefficient retrofit decisions. This study proposes a multi-model probabilistic framework for seismic risk assessment and retrofit planning of electric power systems. The approach integrates: (1) regional seismic hazard characterization with ground motion prediction and spatial correlation models; (2) component-level damage analysis using fragility functions and multi-state damage–functionality mappings; (3) system-level cascading impact evaluation through graph-based island detection and constrained optimal power flow analysis; and (4) retrofit planning via heuristic optimization to minimize expected annual functionality loss (EAFL) under budget constraints. Uncertainty is propagated throughout the framework using Monte Carlo simulation. The methodology is demonstrated on the IEEE 24-bus Reliability Test System, showcasing its ability to capture cascading failures, identify critical components, and generate effective retrofit strategies. Results underscore the framework’s potential as a scalable, data-informed decision-support tool for enhancing the seismic resilience of power infrastructure.

\end{abstract}



\section{Introduction}
\subsection{Motivation}
Electric power networks (EPNs) represent critical infrastructure networks, underpinning essential services such as transportation, healthcare, communication, water supply and emergency response. Their reliable operation is thus essential both during and after the occurrence of natural disasters. Among various natural hazards, earthquakes pose a particularly severe threat to EPNs due to their sudden onset, wide spatial impact, and potential to cause concurrent damage across geographically distributed assets. Past seismic events—such as the 2008 Wenchuan earthquake \cite{qiang2011damage}, the 2011 Christchurch earthquake \cite{kwasinski2014performance}, the 2011 Tohoku earthquake \cite{cimellaro2014physical}, and the 2024 Turkey earthquake \cite{toprak2025aftermath}—have demonstrated the seismic vulnerability of power systems, where damage to generation plants, electric substations, transmission lines, and distribution units led to prolonged blackouts and widespread disruptions across interconnected sectors \cite{guo2017critical,shinozuka2007seismic}. The cascading nature of such failures highlights the interconnectedness of electric power systems and underscores the urgent need for robust seismic risk assessment and mitigation strategies for power infrastructures \cite{azzolin2018electrical}.

It is therefore essential to identify and strengthen vulnerable components through cost-effective and system-informed strategies. Such strategies must account for the spatial correlation of ground motion, network-scale component interdependencies, cascading failure mechanisms, and real-world electrical operational constraints. However, existing research tends to focus on individual components in an EPN, including power plants \cite{xiaohui2020seismic}, electric substations \cite{liang2022seismic}, transmission lines \cite{fu2022seismic}, and distribution circuits, often assuming independent damage behavior across components and binary operability states (i.e., fully operational or failed) \cite{ang1996model}. These simplifications limit the ability to capture network-level risks and operational infeasibility during and after seismic events \cite{cavalieri2014models}. Furthermore, seismic retrofit strategies are frequently derived from local criticality rankings without accounting for system-wide performance \cite{liu2023seismic} or budget constraints \cite{espinoza2020risk}. This calls for a unified, probabilistic, and system-oriented framework that can bridge seismic hazard modeling, fragility assessment, cascading functionality loss, and retrofit planning under uncertainty.

To contextualize these challenges, the following section systematically reviews current research trends and identifies key gaps that motivate the development of the proposed framework.

\subsection{Literature review and research gaps}
Despite increasing attention to the seismic vulnerability of electric power systems, the majority of existing assessments remain confined to the component level. Component-level fragility models have been extensively developed by a wide variety of methods for different power infrastructure elements, including transformers \cite{wen2018seismic,bender2018seismic}, circuit breakers \cite{wen2024improved}, substations \cite{liang2021system,li2019probability}, generation plants \cite{xiaohui2020seismic}, and transmission towers and lines \cite{miguel2021performance,liu2024failure}. These models typically relate ground motion intensity measures (IMs) to the probability of component failure and are useful for estimating localized risk. However, while such component-level fragility assessments are essential, they are insufficient on their own to characterize the overall performance of power systems. Without incorporating system-level interdependencies and cascading effects, these models fail to capture the true operational consequences of seismic events on the network as a whole. Notably, power systems are inherently interconnected, and localized damage can propagate through the network, triggering cascading effects such as power overloads, frequency instability, unintentional islanding, and widespread blackouts \cite{cimellaro2014physical,guo2017critical,azzolin2018electrical}. For instance, Biswas et al. \cite{biswas2020graph} employed graph-theoretical measures to assess network vulnerability under topological disconnection, while Ouyang and Duenas-Osorio \cite{ouyang2014multi} integrated power flow simulation as a tool for evaluating how failures propagate within the power network. Xie et al. \cite{xie2025resilience} further examined network performance degradation using connectivity-based indices under earthquake-induced node failures and line outages. Yet, there is still a lack of fully integrated system-level frameworks that couple physical seismic damage modeling with network operational functionality under realistic conditions.

In addition, a common assumption in literature is that seismic damage to components occurs independently, neglecting the spatial correlation of ground motions \cite{ang1996model,xie2025resilience,pavel2022seismic}. This assumption oversimplifies the nature of earthquake events, in which nearby components often experience similar levels of ground shaking due to correlated seismic IMs (e.g., PGA, Sa) \cite{lee2007uncertainty}. Jayaram and Baker \cite{jayaram2009correlation} and Goda and Hong \cite{goda2008spatial} proposed stochastic simulation methods to generate spatially correlated ground motion fields, and Weatherill et al. \cite{weatherill2015exploring} and Qin et al. \cite{qin2018information} applied these fields in risk assessment of building portfolio and transportation network respectively, highlighting how spatially correlated IMs significantly influence system-wide damage patterns. In the power system domain, Wang et al. \cite{wang2019seismic} examined correlated failure scenarios using Monte Carlo simulation, but integration of spatial correlation into full-scale, functionality-aware risk frameworks remains rare \cite{liang2025probabilistic}. Without considering such correlation, damage distributions may appear unrealistically dispersed, leading to inaccurate loss estimates and misinformed mitigation decisions.

Another critical limitation lies in the use of binary-state assumptions for component performance, following which components are treated as either fully functional or completely failed \cite{ang1996model,wang2019seismic,liang2025probabilistic,lee2025efficient,liu2020resilience}. This binary-state assumption oversimplifies the actual behavior of infrastructure, where partial degradation is more realistic. For example, a partially damaged substation may still operate at derated capacity \cite{liang2023seismic}. Fragility models based on multi-level damage states have been widely adopted in current standards \cite{hamburger2012fema,ulmi2014hazus}, and damage-functionality mapping models that link damage states to partial service levels are gaining attention. However, their application in full-scale seismic risk analysis for power systems remains limited \cite{lagos2019identifying}. Disregarding intermediate functionality states limits the ability to capture cascading failure and quantify realistic residual functionality, potentially resulting in biased risk assessments and suboptimal retrofit strategies.

Even when component-level direct damage is accurately modeled, many previous studies fail to capture system-level functionality accurately, as they often overlook the fundamental electrical constraints governing power flow, such as supply-demand balance, transmission line capacities, generation limits, and power dispatch costs \cite{biswas2020graph,xie2025resilience,salman2017multihazard,xie2022research,sun2019agent}. As a result, these simplified models may yield unrealistic assessments of overall system functionality \cite{ouyang2013comparisons}. More realistic models rely on DC or AC optimal power flow (OPF) formulations \cite{azzolin2018electrical,ouyang2014multi,panteli2017metrics,li2017ac}, which incorporate constraints on generation, demand, and flow capacities, to capture the dynamic nature of power flow rerouting within the EPN after partial damage. But these studies focus on limited contingency scenarios involving the failure of one or two components \cite{da2015method}, whereas earthquakes can simultaneously damage a large number of components to varying degrees, resulting in extensive topological disconnection and frequent OPF convergence failures. Recent studies have attempted to bridge this gap by integrating fragility-based component damage modeling with power flow analysis \cite{pavel2022seismic,wang2019seismic,poulos2017seismic,ferrario2022predictive}. For example, Poulos et al. \cite{poulos2017seismic} coupled seismic fragility sampling with DCOPF to evaluate loss using energy-not-supplied (ENS) in Chile’s power network under probabilistic earthquake scenarios. However, power load demand may also vary significantly in the aftermath of an earthquake due to physical damage, operational disruptions, or shifting consumption patterns, but this aspect is almost always overlooked \cite{blagojevic2022demand,wang2022systematic}. Moreover, few studies account for post-disaster islanding effects \cite{chu2021frequency}, wherein disconnected segments of the EPN may retain partial self-sufficiency depending on local generation and load conditions, rather than becoming entirely non-functional. Without explicitly identifying electrical islands, DCOPF performed on the entire disconnected network can fail to converge or yield infeasible solutions, as disconnected buses prevent a valid global power flow solution. Existing frameworks often fail to integrate these islanded operations and dynamic demand patterns with models of hazard propagation and cascading failures under uncertainty. This omission limits the realism and practical applicability of such approaches in guiding effective retrofit planning.

Seismic retrofit strategies are often based on topological criticality metrics, such as node degree\cite{ferrario2022predictive} and betweenness centrality \cite{chen2023betweenness}. For example, Chen et al. \cite{chen2023betweenness} used betweenness centrality to identify vulnerable network links and guide risk management. While these methods are computationally tractable, they typically overlook how a retrofit at one component may influence system-wide functionality, cascading effects, and interaction between component retrofits. To address this limitation, some researchers have employed local sensitivity analysis approaches, assessing the impact of individual component failures or survivals on overall system performance through extensive simulations to identify critical elements \cite{espinoza2020risk,wang2019seismic,liang2025reliability,liu2021quantifying}[14, 31, 52, 53]. However, such methods remain inherently limited, as they fail to capture the joint impact of multiple retrofitted components. Given the combinatorial nature of retrofit decisions and the interdependence among components, these methods are insufficient for identifying globally optimal solutions. To address this gap, optimization-based methods have gained attention for their ability to explore retrofit combinations under various constraints. Techniques such as integer programming \cite{romero2015seismic}, heuristic algorithms \cite{oboudi2024two}, and reinforcement learning \cite{anwar2024deep}, have shown promise in the optimal retrofit planning context. Nonetheless, many of these approaches remain decoupled from physically grounded power system functionality models or overlook critical factors such as hazard uncertainty. Moreover, few explicitly account for practical constraints, including budget limitations, system-level cascading dynamics, and long-term risk metrics. This highlights the need for a fully integrated, risk-informed optimization framework that can support cost-effective and system-optimal seismic retrofit planning.

\subsection{Objective and contribution}
This study presents an integrated probabilistic framework for seismic risk assessment and retrofit planning of electric power networks, addressing several key limitations in existing methodologies. In particular, the framework targets the following challenges: (1) modeling spatially correlated seismic hazard; (2) capturing multi-level component degradation; (3) evaluating cascading functionality loss under electrical operational constraints; and (4) optimizing retrofit strategies under uncertainty and limited budgets. The main contributions of this work are as follows:
 \begin{itemize}
     \item We integrate regional seismic source, ground motion attenuation, and spatial correlation models to simulate geographically consistent seismic intensity fields. This enables the generation of realistic damage scenarios for large-scale power systems based on the spatial distribution of faults and infrastructure components.
     \item We move beyond binary-state assumptions by implementing multi-level fragility curves and damage–functionality mapping models that capture progressive degradation in component performance, and its impact on cascading failure propagation.
     \item We combine graph-based island detection with DCOPF models to evaluate system functionality under both physical disconnection (i.e., islanding) and operational infeasibility.  Without island detection, global DCOPF analysis often fails on disconnected networks. Analyzing active islands separately ensures feasible and accurate functionality evaluations.
     \item We formulate a risk-informed optimization problem to minimize expected annual functionality loss (EAFL) under budget constraints, leveraging Monte Carlo simulations to capture multi-source uncertainties and using a heuristic genetic algorithm to search for the optimal solution at the system level.     
     \item We validate the proposed framework on the IEEE 24-bus Reliability Test System and further conduct a sensitivity analysis to reveal the relationship between investment levels and risk reduction, providing decision-makers with practical insights into the effectiveness and economic implications of different retrofitting options.     
 \end{itemize}

\section{Methodology}
\begin{figure}
    \centering
    \includegraphics[width=1\linewidth]
    { 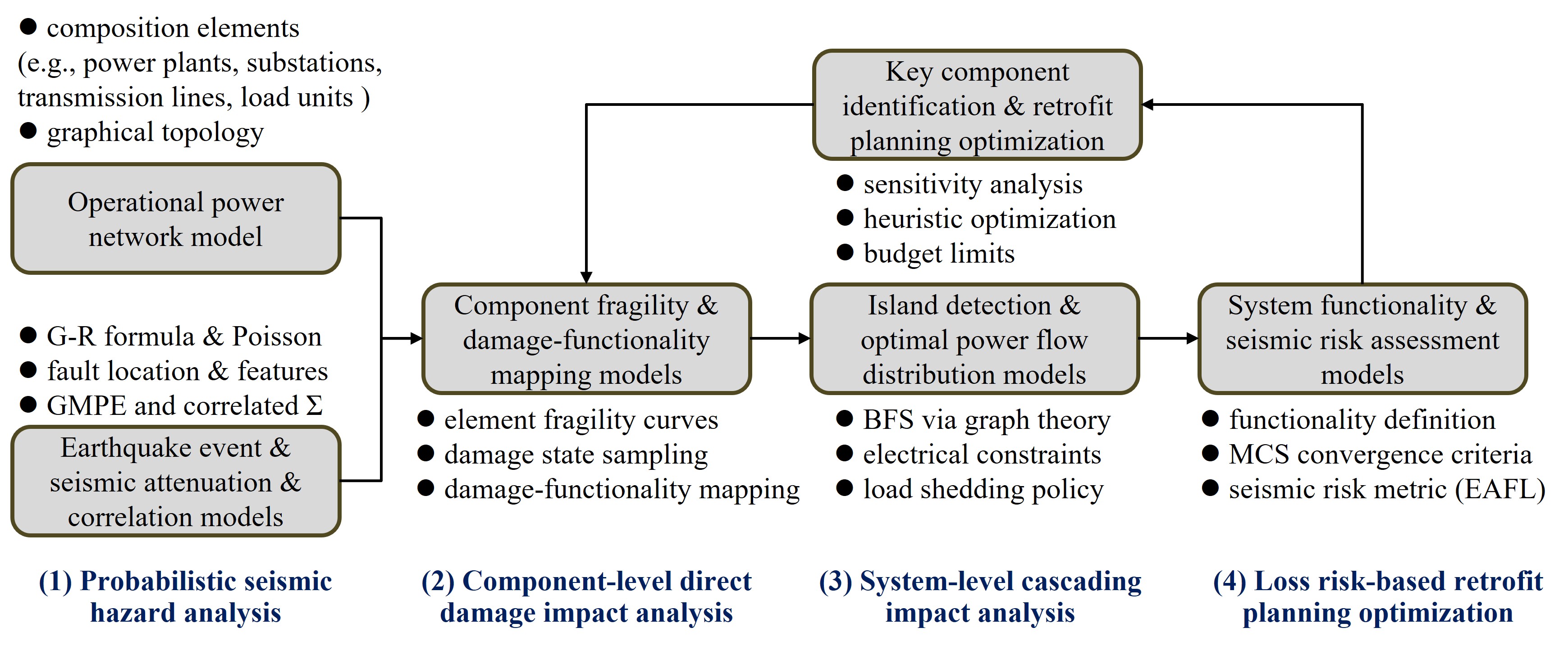}
    \caption{A multi-model probabilistic framework for seismic risk assessment and retrofit planning of EPNs.}
    \label{framework}
\end{figure}

This section outlines the proposed multi-model probabilistic framework for seismic risk assessment and retrofit planning of EPNs. As shown in Figure \ref{framework}, the framework is composed of four interconnected modules: (1) regional seismic hazard modelling with seismic source, attenuation, and correlation models based on the geographic locations of faults and power infrastructure components; (2) component-level direct damage impact analysis leveraging multi-level component fragility curves and damage–functionality mapping models; (3) system-level cascading impact analysis using topology-based island detection and optimal power flow models; and (4) retrofit planning optimization using heuristic search model to minimize seismic functionality loss risk EAFL under budget constraints. These modules are sequentially coupled and form an end-to-end chain from hazard generation to actionable decision-making, with uncertainty propagated throughout the modelling chain via Monte Carlo simulation.

\subsection{Probabilistic seismic hazard analysis with spatial correlation}
Accurate assessment of earthquake impacts on infrastructure systems requires realistic ground motion representation. This module simulates the spatial distribution of ground motion intensity measures (IMs) across the power network, based on stochastic earthquake occurrence, empirical attenuation, and spatial correlation models. The annual frequency of earthquakes exceeding a given magnitude M is estimated using the Gutenberg–Richter (G–R) law \cite{gutenberg1944frequency}:
\begin{equation}
    \log_{10} N(M) = a - b \cdot M
    \label{eq1}
\end{equation}
where $N(M)$ denotes the mean annual rate of earthquakes with magnitude greater than or equal to $M$, and $a$ and $b$ are region-specific seismicity parameters. Assuming a Poisson process, the annual exceedance probability for an earthquake with magnitude $\geq M$ is given by:
\begin{equation}
    P(M) = 1 - e^{-N(M)} = 1 - e^{(-10^{a - b \cdot M})}
    \label{eq2}
\end{equation}
This probabilistic formulation is essential for integrating seismic hazard into system-level risk assessments in later stages of the analysis. Each earthquake scenario is defined by magnitude, epicenter, rupture distance, and fault mechanism parameters. To model the IMs (e.g., peak ground acceleration or spectral acceleration) for each component in the EPN, the Boore–Stewart–Seyhan–Atkinson (BSSA14) ground motion prediction equation (GMPE) is adopted in this study \cite{boore2014nga}. Developed as part of the NGA-West2 project, BSSA14 is widely validated and recommended for active shallow crustal regions and accounts for source characteristics, distance attenuation, and local site effects, as described below:
\begin{equation}
    \ln PGA = F_M + F_D + F_S + \varepsilon
    \label{eq3}
\end{equation}
where $PGA$ = peak ground acceleration (g); $F_M$, $F_P$, and $F_S$ represent functions for magnitude scaling, distance attenuation, and site effects respectively, and detailed formulations can be found in \cite{boore2014nga}; $\varepsilon \sim \mathcal{N}(0, \sigma^2)$ is a zero-mean residual with standard deviation capturing the aleatory uncertainty. To simulate realistic ground motion fields across multiple spatially distributed components, spatial correlation of residuals is introduced following the model proposed by Jayaram and Baker \cite{jayaram2009correlation}, which establishes the correlation coefficient between ground motion residuals at two sites $i$ and $j$ as an exponential decay function of their separation distance and earthquake magnitude:
\begin{equation}
    \rho_{ij} = \exp\left(\frac{-3d_{ij}}{b(M)}\right)
    \label{eq4}
\end{equation}
where $d_{ij}$ represents the inter-site distance (km), and $b(M)$ is a magnitude-dependent correlation length parameter and was empirically calibrated in their study, with representative values ranging from 17.6 km to 31.7 km for magnitudes between 5.0 and 7.0. A linear extrapolation of the fitted trend is used with an upper cap at 40 km in this study to avoid overestimation of spatial correlation at large distances by Equation (\ref{eq5}).
\begin{equation}
  b(M) = \min(5.4 + 4.7 \cdot M, 40)
  \label{eq5}
\end{equation}
Based on this model, the full covariance matrix of residuals is constructed. A zero-mean multivariate normal distribution is then sampled using this covariance matrix to obtain a spatially correlated residual vector. These residuals are then added to the GMPE-predicted logarithmic mean values of PGA to generate spatially consistent samples of $ln(PGA)$. Finally, the resulting samples are exponentiated to obtain spatially correlated PGA fields, which serve as input for fragility-based damage assessment.

\subsection{Component-level direct damage impact analysis}
\begin{figure}
    \centering
    \includegraphics[width=0.5\linewidth]
    { 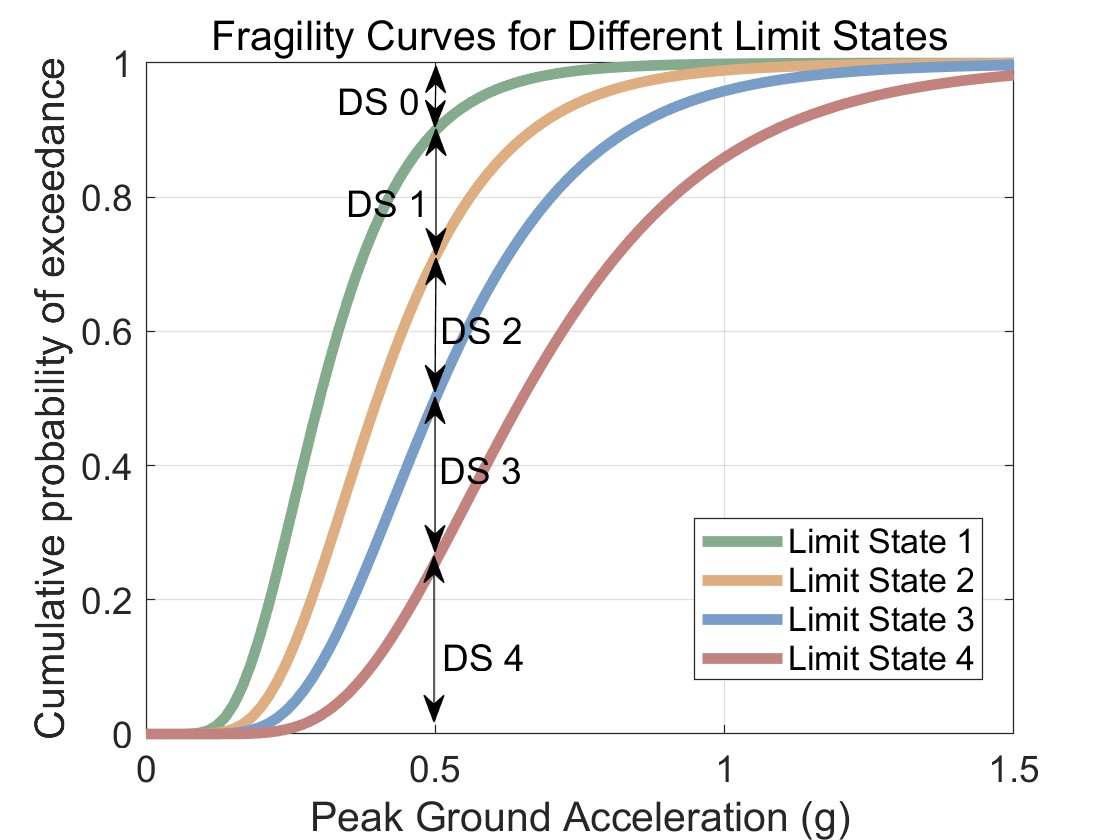}
    \caption{Illustrated fragility curve for a component with five damage states: intact (DS 0), slight damage (DS 1), moderate damage (DS 2), severe damage (DS 3) and complete damage (DS 4). The arrows indicate how a given PGA corresponds to a vertical slice across the curves, and how damage states are probabilistically assigned based on the resulting exceedance values.}
    \label{fragility_curve}
\end{figure}

EPNs comprise diverse components such as generation plants, transmission towers, electrical substations, and load circuits. To assess the direct physical impact of earthquakes on these infrastructure components, this study adopts a probabilistic approach based on component-specific fragility functions. These functions characterize the conditional probability that a component exceeds a certain damage state given the level of ground motion intensity (e.g., PGA). A typical fragility function takes the following lognormal form \cite{liang2022seismic,ulmi2014hazus,liu2021quantifying}:
\begin{equation}
    P(DS \geq ds_k \mid PGA) = \Phi \left( \frac{\ln(PGA) - \ln(\theta_k)}{\beta_k} \right)
    \label{eq6}
\end{equation}
where $P(DS \geq ds_k)$ denotes the probability of exceeding damage state $ds_k$, $\theta_k$ and $\beta_k$ are the median and logarithmic standard deviation parameters of the fragility curve corresponding to damage state $k$, and $\Phi(\cdot)$ is the standard normal cumulative distribution function. Multiple fragility curves are defined for each component to represent different limit states of increasing severity (e.g., Slight, Moderate, Extensive, Complete), as illustrated in Figure \ref{fragility_curve}.

\begin{figure}
    \centering
    \includegraphics[width=0.5\linewidth]
    { 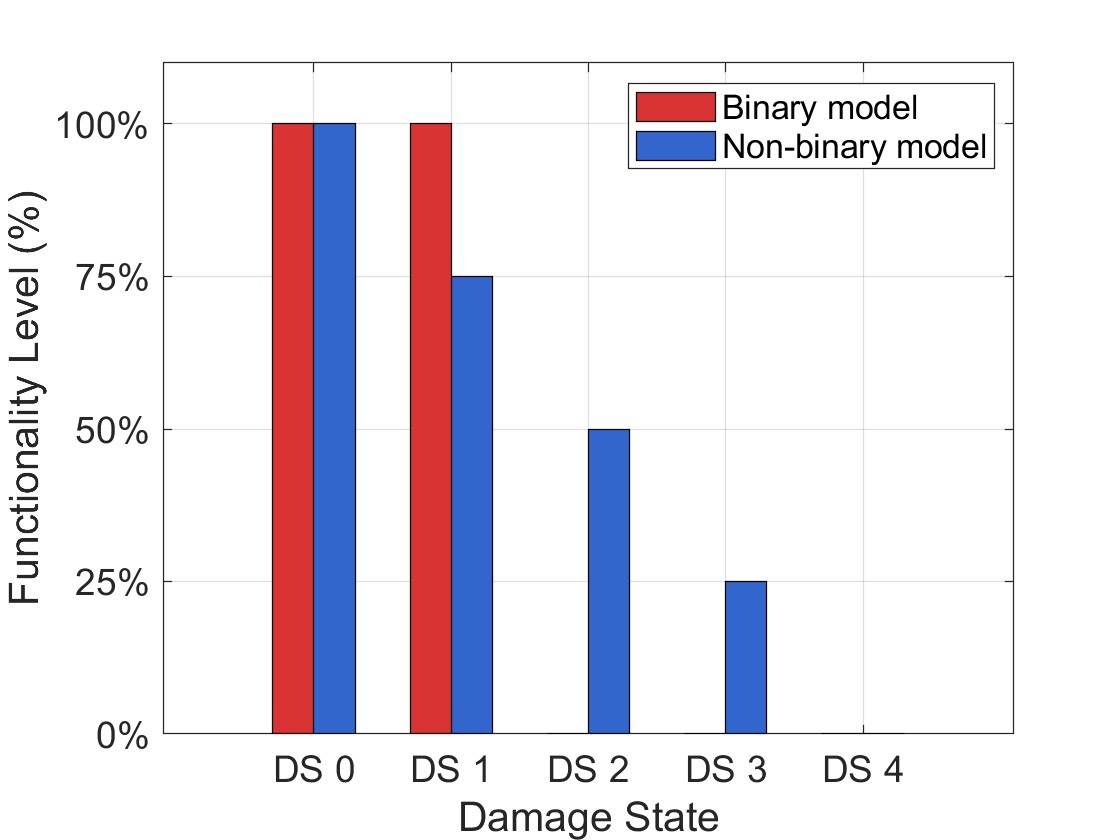}
    \caption{Mapping from damage states to functionality levels for diverse types of power system components. The binary model applies to bus nodes, while the non-binary model describes progressive functionality loss in generators, loads, and substations.}
    \label{Mapping_damage_to_functionality}
\end{figure}

For each ground motion realization generated in the hazard module, the PGA value at each component location is used as input to its corresponding fragility curves, and a Monte Carlo-based inverse sampling scheme is employed to determine the discrete damage state. Specifically, the exceedance probabilities corresponding to the given PGA value are computed from the fragility curves of different limit states. A random number is then drawn from a uniform distribution $u \sim \mathcal{U}(0, 1)$, and the damage state is assigned based on the interval in which $u$ falls between two successive exceedance probabilities. This sampling logic is mathematically expressed by Equation (7) (let $P(ds_0)=1$ and$ P(ds_5)=0$ by convention) and illustrated in Figure \ref{fragility_curve}, where a vertical slice at the given PGA intersects the fragility curves, defining probability ranges for damage states DS0 through DS4.
\begin{equation}
    DS = d_{s_k}, \text{ if } P(d_{s_{k-1}}) < u \leq P(d_{s_k})
    \label{eq7}
\end{equation}
Once the damage state is determined, it is translated into the component’s corresponding functionality levels through predefined damage–functionality relationships. This mapping reflects the component's residual ability to perform its intended role under degraded physical conditions and differs by component type. As illustrated in Figure \ref{Mapping_damage_to_functionality}, two types of relationships are considered in this study:

a) Binary model is used for bus nodes, which act solely as network connection points. They either enable connectivity or become isolated due to failure. Therefore, they are assumed to remain fully functional (100\%) in undamaged and minor damaged states (DS0–DS1) and become entirely non-functional (0\%) once moderate or severe damage (DS2 and above) occurs. 

b) Non-binary models are applied to generation plants, load units, and power substations, where intermediate damage states result in partial operability (e.g., reduced generation capacity, partial load loss, or derated transformation performance). For instance, components may operate at 75\%, 50\%, or 25\% of their nominal capacity under increasing damage levels (DS1–DS3), before complete failure (DS4).

This differentiated modeling strategy captures the distinct physical roles and operational behaviors of various component types, allowing for more realistic system-level cascading analysis and functionality assessment in subsequent steps.

\subsection{System-level cascading impact analysis}
\begin{figure}
    \centering
    \includegraphics[width=0.4\linewidth]
    { 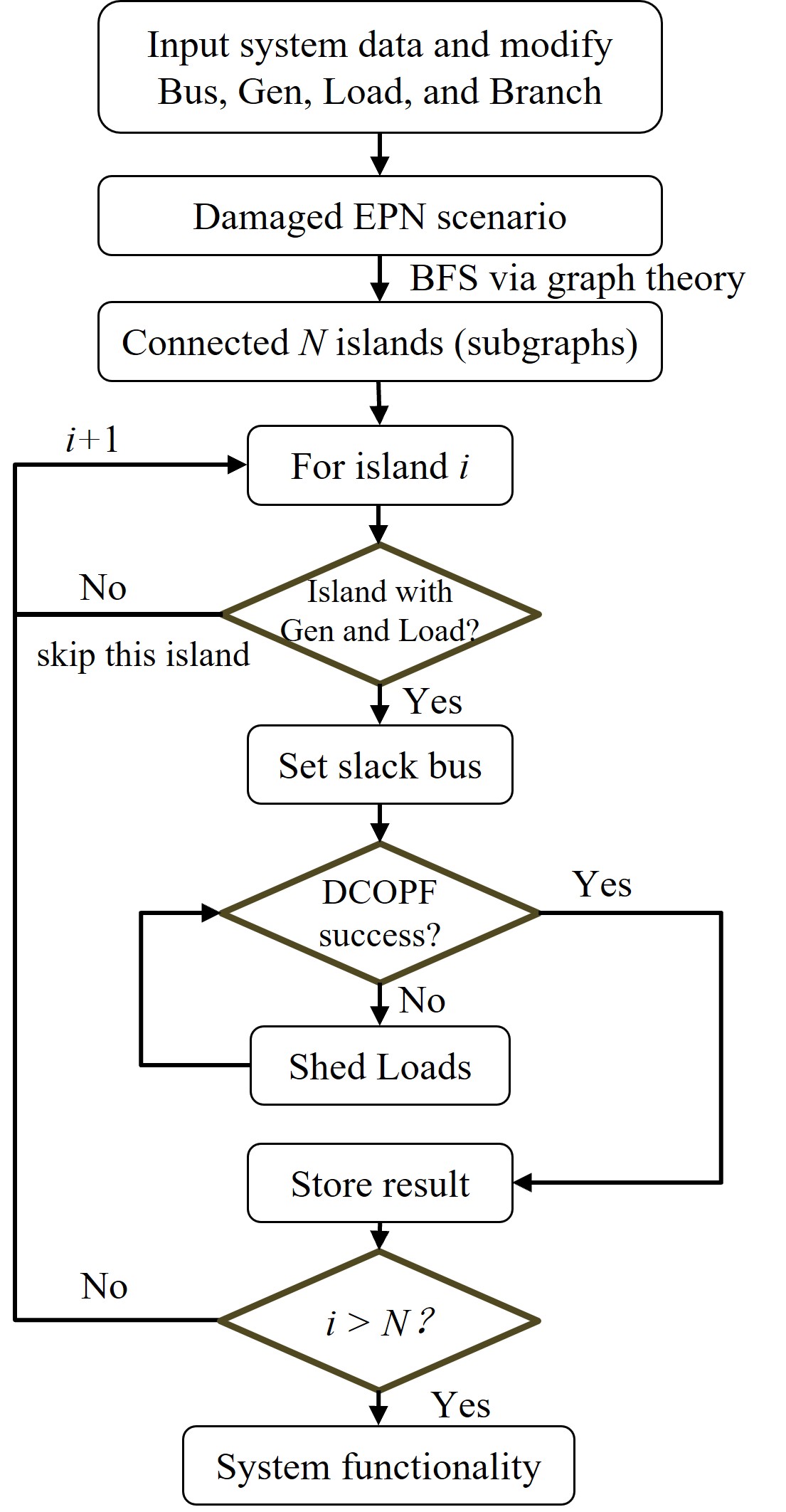}
    \caption{Workflow for system-level cascading impact analysis. The process consists of two sequential steps: (i) identifying electrically isolated subnetworks (islands) through graph-based topological analysis, and (ii) evaluating power redistribution within each island using DCOPF.}
    \label{flowchart}
\end{figure}

Beyond component damage, EPN performance depends on system topology and operational feasibility. This module evaluates both physical disconnection and power redistribution under operational constraints, capturing the cascading effects that amplify localized failures. The procedure involves two sequential steps: (i) island identification through graph-based topological analysis and (ii) power flow rerouting via DC optimal power flow (DCOPF), as shown in Figure \ref{flowchart}.

We first represent the power network as a graph $\mathcal{G} =(V, E)$, where nodes correspond to buses, generation plants, and load units, and edges represent transmission lines or substations. Based on the sampled post-earthquake damage states, all failed nodes and disconnected edges are removed from the network to reflect post-earthquake connectivity status. An updated binary adjacency matrix $\mathbf{A} \in \{0,1\}^{n \times n}$ is then constructed, capturing the post-event connectivity. Specifically, a connection between two nodes is retained only if both terminal buses are operational and the corresponding transmission line remains functional. Mathematically, the adjacency matrix is defined as:
\begin{equation}
    A_{ij} = 
    \begin{cases} 
    1 & \text{if buses } i \text{ and } j \text{ and line } ij \text{ is intact} \\
    0 & \text{otherwise}
    \end{cases}
    \label{eq8}
\end{equation}
To identify electrically isolated subnetworks (i.e., electrical islands), a graph traversal algorithm based on breadth-first search (BFS) \cite{cormen2022introduction} is applied to the adjacency matrix A. Starting from each unvisited node, BFS explores all connected nodes by expanding outward layer by layer, grouping them into the same subgraph. Repeating this process partitions the network into isolated subgraphs $\mathcal{G}_1$, $\mathcal{G}_2$, ..., $\mathcal{G}_N$, each representing a distinct island. This island detection enables independent evaluation of each subnetwork, accounting for localized generation, load, and connectivity in post-event system functionality assessment.

For an island $\mathcal{G}_i$ to be considered operationally viable, it must (i) contain at least one operational generator and (ii) include at least one load bus with nonzero demand. Otherwise, the island is deemed inoperable and excluded from further computation. If the island lacks a slack bus, a generator-connected bus is designated as the slack to enable power flow solvability. Once verified, a DCOPF formulation problem is solved for each viable island using MATPOWER’s built-in interior-point solver (MIPS) \cite{zimmerman2010matpower}, which minimizes generation cost subject to nodal power balance and line flow constraints under degraded capacity conditions. The optimization problem can be expressed as:
\begin{equation}
\begin{aligned}
    & \min_{PG, \theta} \sum_{i \in G} C_i(PG_i) \\
    & \text{subject to} \\
    & \begin{cases} 
      PG_i - PD_i = \sum_{j \in N(i)} B_{ij}(\theta_i - \theta_j) \quad \text{(Nodal balance constraint)} \\ 
      |B_{ij}(\theta_i - \theta_j)| \leq \alpha_{ij} \cdot P_{\text{RATE},ij} \quad \text{(Line flow capacity constraint)} \\ 
      \alpha_i \cdot PG_i^{\text{min}} \leq PG_i \leq \alpha_i \cdot PG_i^{\text{max}} \quad \text{(Generator constraint)} 
    \end{cases}
\end{aligned}
\label{eq9}
\end{equation}
Here, $PG_i$ and $C_i$ are the power output and the cost coefficient of generator $i$, respectively; $PD_i$ and $\theta_i$ are the power demand and voltage angle at bus $i$, respectively; $B_{ij}$ and $P_{RATE,ij}$ are the admittance (reciprocal of reactance) and transmission limit of line $(i, j)$, respectively; $PG_i^{\text{min}}$ and $PG_i^{\text{max}}$ are the minimum and maximum output limits of generators, respectively; and $\alpha_{ij} \in [0,1]$ and $\alpha_i \in [0,1]$ denote the residual capacity ratio of line $(i, j)$ and node component $i$ based on their damage states and functionality mappings, respectively.

If the DCOPF does not converge, a retry mechanism is implemented where the smallest non-zero load in this island is sequentially shed until convergence or the retry limit is reached. This reflects emergency operational practice in real-world post-earthquake conditions. Once converged, we record the total amount of load served (in MW) in that island, denoted as:
\begin{equation}
    F_{\mathcal{G}_i}^{served} = 
    \begin{cases} 
        \sum_{j \in L} PD_{j}^{served} & \text{if DCOPF converges} \\
        0 & \text{otherwise}
    \end{cases}
    \label{eq10}
\end{equation}
where $PD_j^{served}$ is the actual load met at load bus $j$ after generation-dispatch adjustment and potential load shedding. The system-level functionality for a damaged EPN scenario $k$ is computed as the aggregate sum across all islands:
\begin{equation}
    F_k = \sum_{i=1}^{N} F_{\mathcal{G}_i}^{\text{served}}
    \label{eq11}
\end{equation}
This two-stage process accounts for the physical disconnection, degraded component capacities, and dynamic load reallocation, offering a more realistic picture of system-level resilience compared to traditional fragility-based assessments or topology-only models.

\subsection{Loss risk-based retrofit planning optimization}
To optimize retrofit strategies for EPNs, it is essential to first quantify the system-level risk induced by earthquakes. The expected annual functionality loss (EAFL) is introduced as a comprehensive risk metric by integrating the severity and frequency of functionality loss across a range of seismic intensities, providing a scalar measure for risk-based comparison across retrofitting alternatives under uncertainty: 
\begin{equation}
    EAFL = \int_{0}^{\infty} [1 - \overline{F}_N(M)] \cdot \lambda(M) \, d(M) \approx \sum_{i=1}^{I} \lambda(M_i) \cdot [1 - \overline{F}_N(M_i)]
    \label{eq12}
\end{equation}
where $\lambda(M)$ is the annual occurrence rate of earthquake with a magnitude of $M$, which can be derived from Equation (\ref{eq2}) in section 2.1 on probabilistic seismic hazard analysis; $\overline{F}_N(M)$ is the average normalized system functionality given that specific $M$, which can be obtained by Equation (\ref{eq11}) based on the results from Monte Carlo simulations (MCS).
\begin{equation}
    \overline{F}_N = \frac{1}{K} \sum_{k=1}^{K} \frac{F_k}{F_o}
    \label{eq13}
\end{equation}
where $F_k$ represents the system-level functionality for a damaged scenario $k$; $F_o$ is the pre-earthquake system baseline functionality; $K$ is the number of simulated realizations per magnitude. To ensure the reliability of estimated functionality statistics, convergence checks are applied during the MCS process using (i) sample mean stability and (ii) confidence interval width.
\begin{equation}
    \frac{|\hat{\mu}_n - \hat{\mu}_{n-1}|}{\hat{\mu}_{n-1}} < \tau
    \label{eq14}
\end{equation}
where $\hat{\mu}_n$ is the sample mean functionality and $\tau$ is a small threshold (e.g., 1\%).
\begin{equation}
    \text{Width}_{95\%} = 2 \cdot z \cdot \frac{\hat{\sigma}_n}{\sqrt{n}} < \delta
    \label{eq15}
\end{equation}
where $\hat{\sigma}_n$ is the sample standard deviation; $n$ is the number of simulation samples; $z$ is the standard normal critical value (i.e., 1.96 for 95\% confidence); and $\delta$ is a tolerance (e.g., 5\%).

Building upon this risk metric, we then formulate a risk-based retrofit planning optimization framework that aims to minimize EAFL through targeted upgrades of vulnerable components. Let the power network consist of $n$ candidate components (e.g., substations, generators, load units), each associated with a binary decision variable $x_i$, which may be altered through retrofit actions. The decision vector is defined as:
\begin{equation}
    \boldsymbol{x} = [x_1, x_2, \ldots, x_n], \; x_i \in \{0, 1\}
    \label{eq16}
\end{equation}
where $x_i =1$ denotes that component $i$ is retrofitted and $x_i =0$ otherwise. Let $c_i$ be the cost of retrofitting component $i$, and $B$ the total available budget, then the constrained optimization problem can be expressed as:
\begin{equation}
    \begin{aligned}
        & \min_{x \in \{0,1\}^n} \, EAFL(x), \\
        & \text{subject to} \, \sum_{i=1}^{n} c_i x_i \leq B
    \end{aligned}
    \label{eq17}
\end{equation}
To guide the search for effective solutions in this combinatorial space, a one-at-a-time (OAT) sensitivity analysis is first conducted to identify the most influential components. For each component $i$, its median fragility parameter $\theta_i$ is perturbed individually (e.g., upgraded or downgraded), while holding others fixed, and the change in EAFL is recorded. The sensitivity index $S_i$ for component $i$ is calculated as:
\begin{equation}
    S_i = f(\theta_1^0, \ldots, \theta_i^{pert}, \ldots, \theta_n^0) - f(\theta_1^0, \ldots, \theta_i^0, \ldots, \theta_n^0)
    \label{eq18}
\end{equation}
where $\theta_i^{pert}$ denotes a perturbed fragility value and $f(\cdot)$ denotes the EAFL under the specified configuration. Components with large absolute values of $S_i$ are prioritized as key candidates for retrofitting in the optimization stage.

Given the non-convexity and high dimensionality of the problem, a heuristic optimization algorithm, such as a genetic algorithm (GA), is employed to explore the solution space of retrofit strategies as follows.

1)  Each candidate solution is encoded as a binary retrofit vector $\boldsymbol{x} = [x_1, x_2, \ldots, x_n]$.

2) An initial population is generated by combining sensitivity-informed seeding on critical components with random sampling to accelerate convergence while maintaining diversity.

3) For each individual candidate, fragility parameters are updated for retrofitted components accordingly, followed by Monte Carlo simulations to estimate EAFL. A penalized fitness function is applied to enforce budget constraints:
\begin{equation}
    fitness(\boldsymbol{x}) = EAFL(\boldsymbol{x}) + \gamma \cdot \max \left( 0, \sum_{i=1}^{n} c_i x_i - B \right)
    \label{eq19}
\end{equation}
where $\gamma$ is a penalty coefficient.

4) The population is refined iteratively through selection, crossover, and mutation. Top-performing individuals based on fitness rankings are retained unchanged to the next generation (elitism), ensuring solution quality does not degrade.

5) The optimization process terminates when the maximum number of generations is reached or when the best fitness value shows convergence over successive generations. The output is an optimal retrofit configuration that yields the lowest EAFL given a specific budget.

This integrated framework enables both critical component identification and cost-effective retrofit planning, capturing the complex interplay between hazard uncertainty, component fragility, cascading failures, and potential retrofit benefits in a probabilistic seismic environment. Furthermore, a budget–risk sensitivity analysis can be conducted based on this framework to capture the relationship between investment and risk reduction, which offers decision-makers practical guidance on balancing resilience gains against economic constraints, as will be demonstrated in the following case study.

\section{Case Study}
\subsection{Power network configuration}
To demonstrate the proposed framework, we conduct a comprehensive case study on the IEEE 24-bus Reliability Test System \cite{subcommittee1979ieee}, a widely adopted benchmark in power system reliability research \cite{cheng2021reliability}. This system provides a well-structured yet non-trivial network topology that captures essential operational characteristics of real-world power systems, including generation-load balance, inter-area power exchange, and vulnerability to topological fragmentation. Its moderate scale enables rigorous simulation of cascading impacts and retrofitting optimization under various seismic scenarios, while maintaining computational tractability.

\begin{figure}
    \centering
    \includegraphics[width=0.95\linewidth]
    { 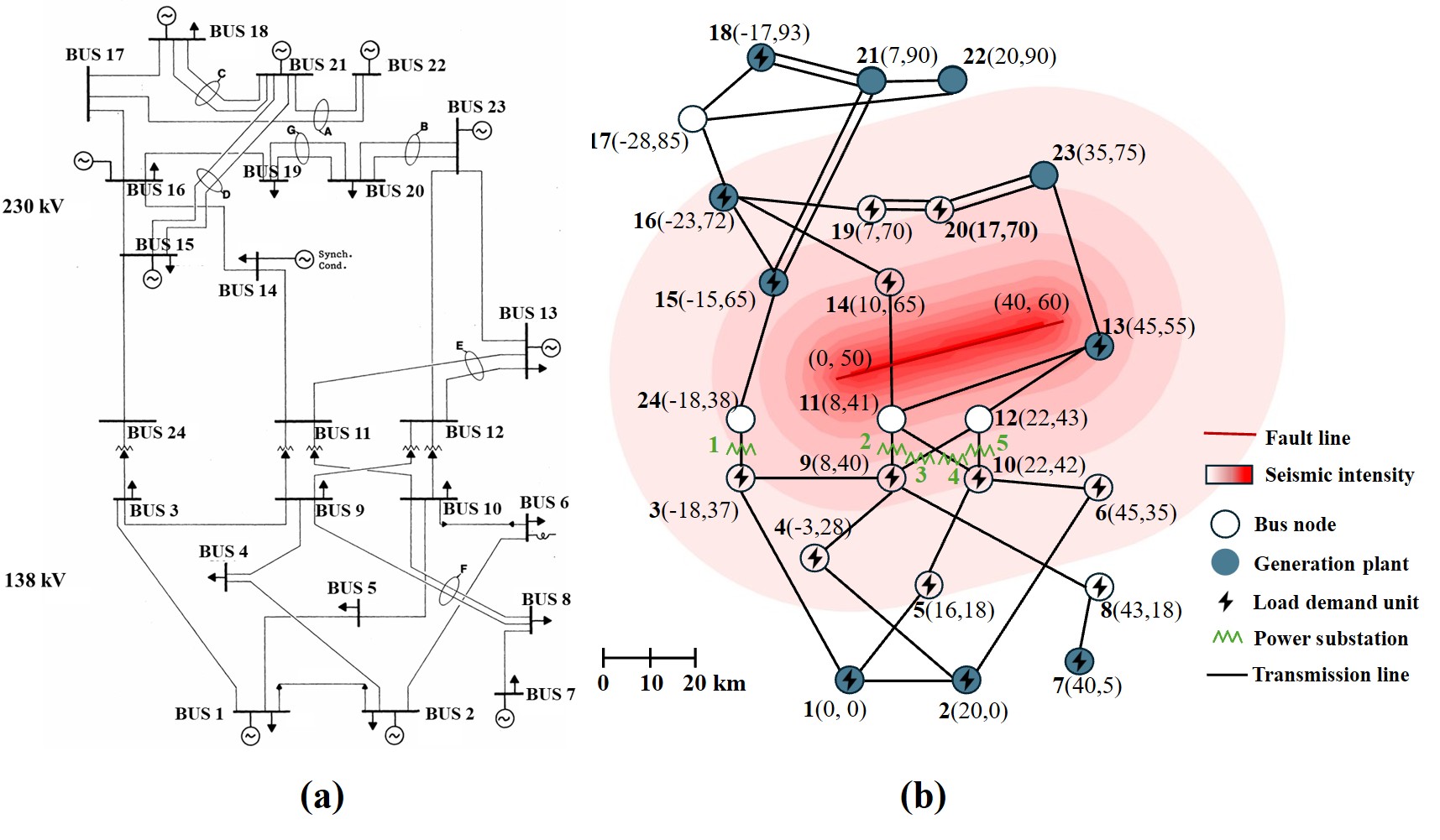}
    \caption{Illustrated IEEE 24-bus Reliability Test System (a) one-line diagram \cite{subcommittee1979ieee}; (b) abstracted network and component layout diagram subjected to an earthquake scenario with a fault line (dashed red) defined by endpoints at (0,50) and (40,60). The red shaded contours represent the seismic intensity field, with darker zones indicating higher seismic intensity. Generation plants (blue circle), load units (lightning symbol), and substations (green polyline) are all interconnected through bus nodes, which serve as the core junction points for power injection, withdrawal, and routing within the network. The geographic coordinates of each bus node are indicated in parentheses. The black lines represent the transmission lines connecting the bus nodes.}
    \label{IEEE-24-bus}
\end{figure}

As shown in Figure \ref{IEEE-24-bus}(a), the illustrated system is organized into two voltage levels: 138 kV and 230 kV and consists of 24 buses that are linked by 38 transmission lines. Bus types include generation-only, load-only, and mixed-function nodes, and each bus is associated with operational attributes such as voltage angle, generation capacity, and load demand. This one-line diagram can be abstracted into an undirected network graph composed of nodes and edges representing different infrastructure components, as shown in Figure \ref{IEEE-24-bus}(b), the system consists of 10 generation plants (blue circle), 17 load units (lightning symbol), and 5 power substations (green polyline), and they are all interconnected through bus nodes, which serve as the core junction points for power injection, isolation, and routing within the network. The pre-earthquake system baseline functionality is 2850 MW. To simulate spatially distributed seismic impact, each bus node is assigned a 2D coordinate and a hypothetical fault line is defined crossing the central and northeast regions of the system, with endpoints located at (0, 50) and (40, 60). For illustrative purposes and without loss of generality, the power system is assumed to be located in southwestern China with a representative site condition of $V_{S30}=760$ m/s, and a strike-slip fault mechanism with a moment magnitude of 8.0 is assumed to characterize the earthquake scenario. This scenario-based setup enables us to assess the propagation of seismic damage through physical disconnection, power flow infeasibility, and service loss, and to test the effectiveness of retrofit strategies in mitigating risk under realistic spatially distributed earthquake events. Given this scenario, the parameters required for the seismic source model (Equation (\ref{eq1})) and the BSSA14 GMPE (Equation (\ref{eq3})) can be specified by consulting relevant literature. The detailed formulations and adopted coefficients are provided in Appendix A for transparency and reproducibility. Accordingly, stochastic seismic intensity fields in terms of PGA with spatial correlation can be generated following the description in section 2.1, as depicted schematically in Figure \ref{IEEE-24-bus}(b) by the red shaded contours.

The seismic fragility parameters for buses, generation plant, load units, and substations follow the Hazus Earthquake Model developed by FEMA \cite{ulmi2014hazus}. These include median capacity thresholds and dispersion values with respect to PGA, as summarized in Table \ref{tab:1}. It is worth noting that the Hazus model does not provide fragility data for transmission towers or overhead lines. This omission reflects a broader assumption in existing studies: transmission structures are generally robust against ground shaking due to their design for extreme loads such as wind, ice, and cascading tower failure. Consequently, in this study, transmission towers are assumed to remain functional under earthquake excitation. This simplification is supported by prior experimental tests \cite{liang2020shaking} and field evidence \cite{salman2017multihazard} indicating that seismic damage to such structures more often arises from secondary geotechnical failures (e.g., landslides or liquefaction), for which data remain limited.

\begin{table}[h]
    \centering
    \caption{Seismic fragility parameters for different EPN components}
    \begin{tabular}{lcccccccc}
        \toprule
        \textbf{Damage} & \multicolumn{2}{c}{\textbf{Bus node}} & \multicolumn{2}{c}{\textbf{Generation plant}} & \multicolumn{2}{c}{\textbf{Load unit}} & \multicolumn{2}{c}{\textbf{Substation}} \\
        \textbf{state} & $\theta(g)$ & $\beta$ & $\theta(g)$ & $\beta$ & $\theta(g)$ & $\beta$ & $\theta(g)$ & $\beta$ \\
        \midrule
        Slight & 0.13 & 0.65 & 0.10 & 0.60 & 0.24 & 0.25 & 0.10 & 0.60 \\
        Moderate & 0.26 & 0.50 & 0.22 & 0.55 & 0.32 & 0.23 & 0.20 & 0.50 \\
        Extensive & 0.34 & 0.40 & 0.49 & 0.50 & 0.58 & 0.15 & 0.30 & 0.40 \\
        Complete & 0.74 & 0.40 & 0.79 & 0.50 & 0.89 & 0.15 & 0.50 & 0.40 \\
        \bottomrule
    \end{tabular}
    \label{tab:1}
\end{table}

\section{Results and Discussions}
\subsection{Scenario-based damage and cascading impact}
\begin{figure}
    \centering
    \includegraphics[width=0.95\linewidth]
    { 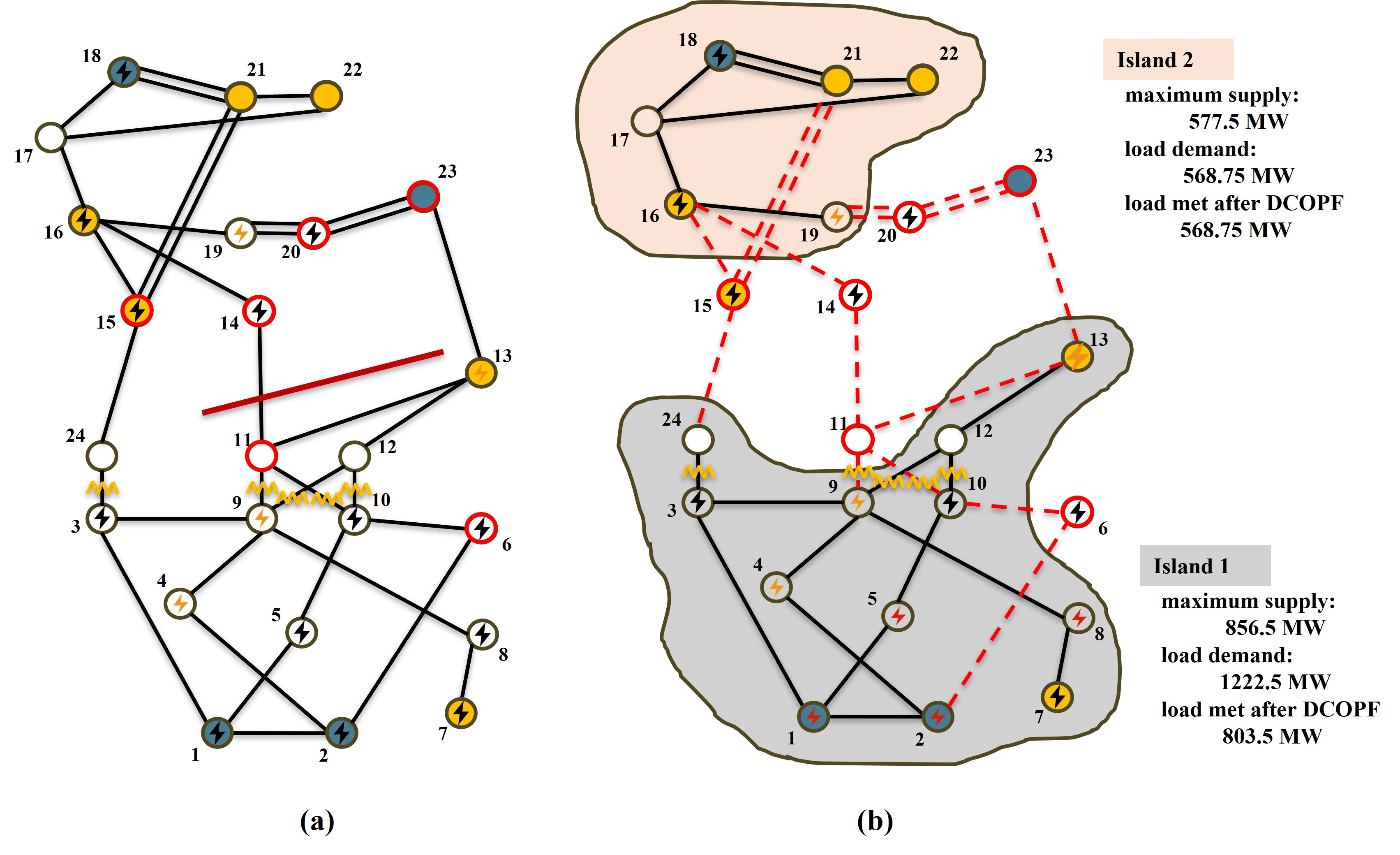}
    \caption{(a) A sampled damage state realization for the EPN components after earthquake (components in red are completely damaged and must be isolated from the power network, while components in orange represent intermediate damage states with partial functionality); (b) post-damage network topology with two identified electrical islands and DCOPF distribution results for the scenario.}
    \label{Sampled_scenario}
\end{figure}

Based on the generated spatially correlated PGAs and the listed seismic vulnerability parameters in section 3, the damage states of all the involved EPN components can be determined by a Monte Carlo-based inverse sampling scheme following Equations (\ref{eq6})(\ref{eq7}) and the description in section 2.2. For better illustration, one sample realization of fragility-based damage states is visualized in Figure \ref{Sampled_scenario}(a), where components shown in red represent those that have experienced complete damage and must be isolated from the power system; components highlighted in orange indicate intermediate damage states and are still able to retain partial operational capacity, with functionality levels determined by their assigned damage states and the predefined damage–functionality mapping models in Section 2.2.

Following the component-level damage realization, the network topology is updated by removing failed nodes and links, and electrically isolated subnetworks (i.e., islands) are identified using the graph-based BFS algorithm. As depicted in Figure \ref{Sampled_scenario}(b), two active islands are detected with non-zero generation capacity and load demand. These islands are evaluated independently using DCOPF-based dispatch to determine the residual functionality, with load shedding implemented iteratively until power balance is achieved within each island, as detailed in section 2.3. The results are also provided in Figure \ref{Sampled_scenario}(b): for Island 1 consisting of buses 16-17-18-19-21-21-22, a total of 803.5 MW of load demand is successfully met out of 1222.5 MW total demand, under a maximum supply capacity of 856.5 MW, reflecting partial load shedding due to electrical constraints (Load 5, 8, 2, 1 were shed sequentially sorted by their PDs from lowest to highest); for island 2 consisting of buses 1-2-3-4-5-7-8-9-10-12-24, DCOPF results show that 568.75 MW of the load demand is fully served given 577.5 MW of available power supply. As a result, the system-level functionality under this damage scenario is computed via Equation (11) as the sum of the load served across all the islands as 1372.25 MW, corresponding to a post-event system functionality level of approximately 54.5 \%.

\subsection{Probability-based functionality loss risk}
\begin{figure}
    \centering
    \includegraphics[width=1\linewidth]
    { 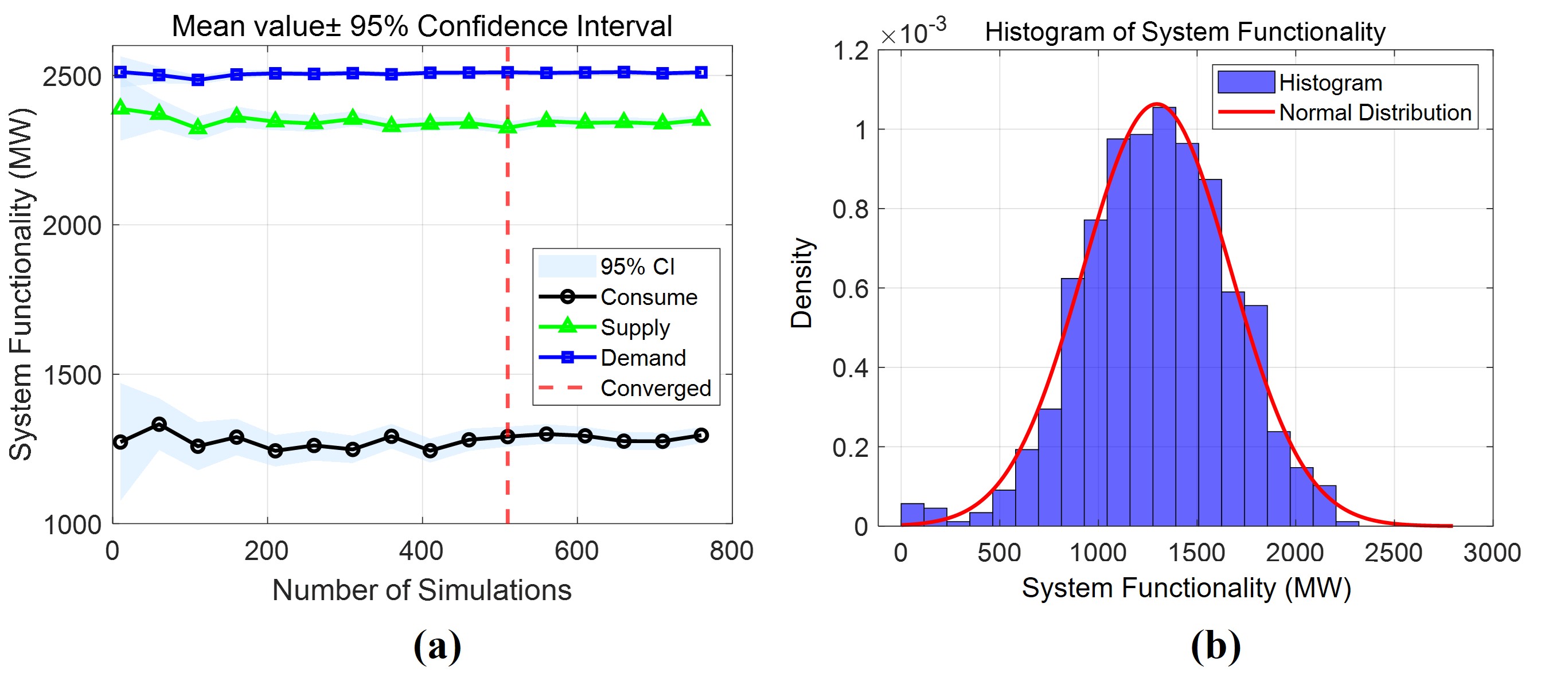}
    \caption{(a) Convergence behavior of Monte Carlo simulations under the seismic magnitude of 8.0 showing sample mean functionality and 95\% confidence bounds; (b) histogram and normal fit of post-earthquake system functionality distribution across simulations.}
    \label{Convergence_behavior}
\end{figure}

To capture uncertainty in both ground motion intensity and fragility-based damage modelling, a MCS is conducted under the fixed earthquake magnitude of 8.0. Within each simulation, spatially correlated ground motions are generated, component damage is sampled, and system functionality is evaluated accordingly. The simulation is iterated until both convergence criteria—sample mean stability and 95\% confidence interval width (Equations (\ref{eq14}) and (\ref{eq15}))—are satisfied. As shown in Figure \ref{Convergence_behavior}(a), the simulation converges after approximately 500 simulations, yielding a final mean functionality estimate of 1262.5 MW, (44.3\% of the total pre-event demand). Notably, the post-damage total supply capacity and total load demand declined to 73.7\% and 88.1\% of their original values, respectively. Despite sufficient residual generation capacity, the actual functionality remains substantially lower, underscoring that the functionality loss is not solely due to physical component-level damage but also driven by system-level topological disconnection and power flow infeasibility. These combined effects reduce the system's ability to maintain power delivery, even when residual generation capacity is available. This finding reinforces the necessity of integrating DCOPF-based analysis, which captures realistic dispatch constraints and interdependencies that are often overlooked in topology-only or capacity-based evaluations, to enable accurate assessment of post-earthquake system performance. The histogram of the system functionality obtained across all Monte Carlo samples are presented in Figure \ref{Convergence_behavior}(b), which approximates a normal distribution, highlighting the stochastic nature of functionality loss driven by random damage realizations and cascading effects. This distribution provides a more comprehensive understanding of expected system performance under the given seismic scenario and enables a direct quantification of the uncertainty range with a probabilistic perspective.

\begin{figure}
    \centering
    \includegraphics[width=0.5\linewidth]
    { 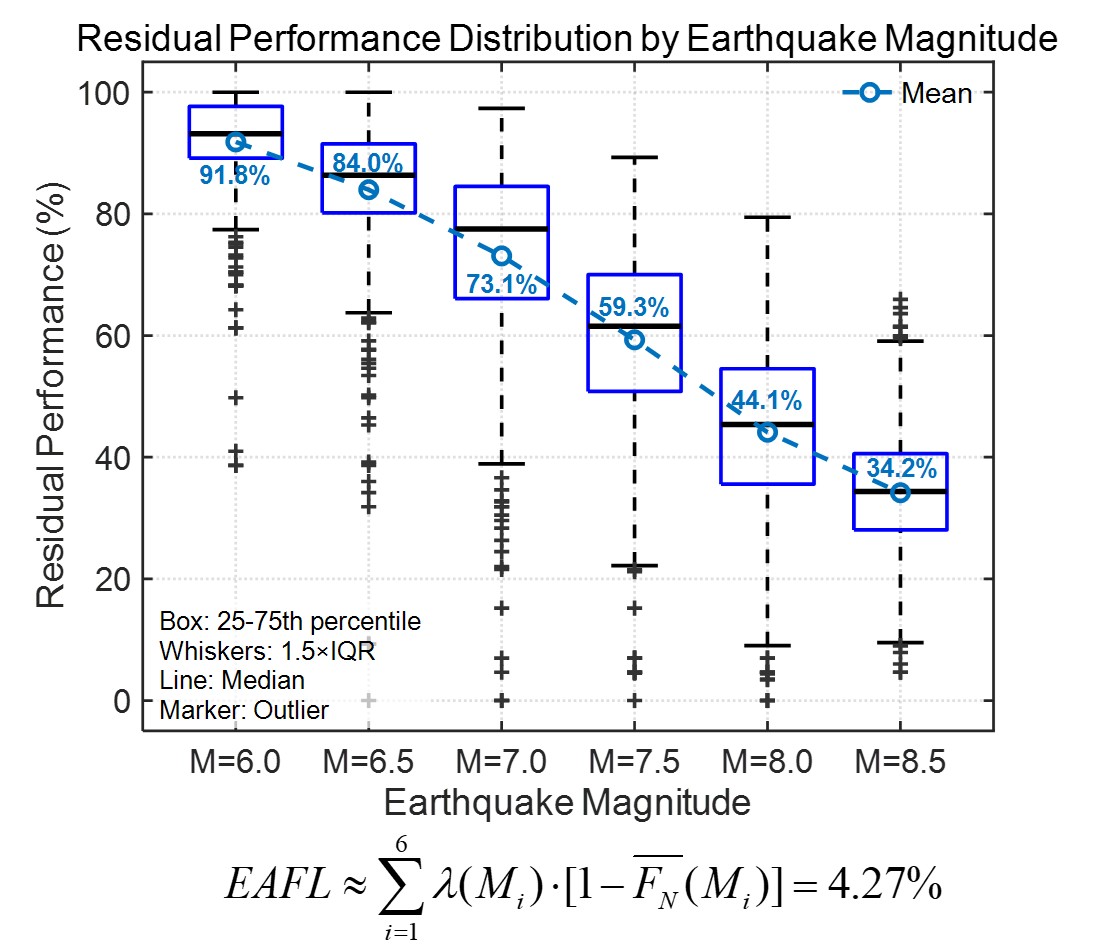}
    \caption{Boxplot of residual system functionality distributions across earthquake magnitudes. Boxes represent the interquartile range (25th–75th percentile), whiskers extend to 1.5×IQR, and blue circles denote mean normalized functionality values used in EAFL computation}
    \label{Boxplot}
\end{figure}

A comprehensive seismic risk assessment is conducted by extending the analysis over a spectrum of magnitudes ranging from 6.0 to 8.5, each weighted by its annual occurrence rate from the probabilistic hazard model described in Section 2.1. For each magnitude level $M_i$, Monte Carlo simulations are conducted to estimate the average normalized system functionality $ \overline{F}_N$ through Equation (\ref{eq13}), following the same procedure outlined in Section 4.2. The results are summarized in Figure \ref{Boxplot}, which presents the distribution of residual system performance across all simulations in the form of boxplots. Each box represents the interquartile range (25th to 75th percentile), whiskers extend to 1.5 times the interquartile range, and outliers are shown as individual markers. The mean functionality is denoted by the blue circle, with exact values annotated for each magnitude level. As expected, a clear downward trend is observed due to the compound impact of structural damage, network disconnection, and operational limitations as earthquake intensity increases. Specifically, the average functionality drops from 91.8\% at M=6.0 to 34.2\% at M=8.5, with increased variability observed at higher intensities. These magnitude-dependent functionality estimates serve as direct input to the seismic risk integration in Equation (\ref{eq12}), resulting in an EAFL of 4.27\% that indicates a moderate yet non-negligible seismic risk to the current power system configuration, as shown in Figure \ref{Boxplot}. This quantification metric forms the basis for the subsequent risk-informed retrofit strategy optimization in the next section, which aims to minimize expected losses under budget constraints.

\subsection{Risk-based critical component identification and retrofit optimization}
\begin{figure}
    \centering
    \includegraphics[width=0.75\linewidth]
    { 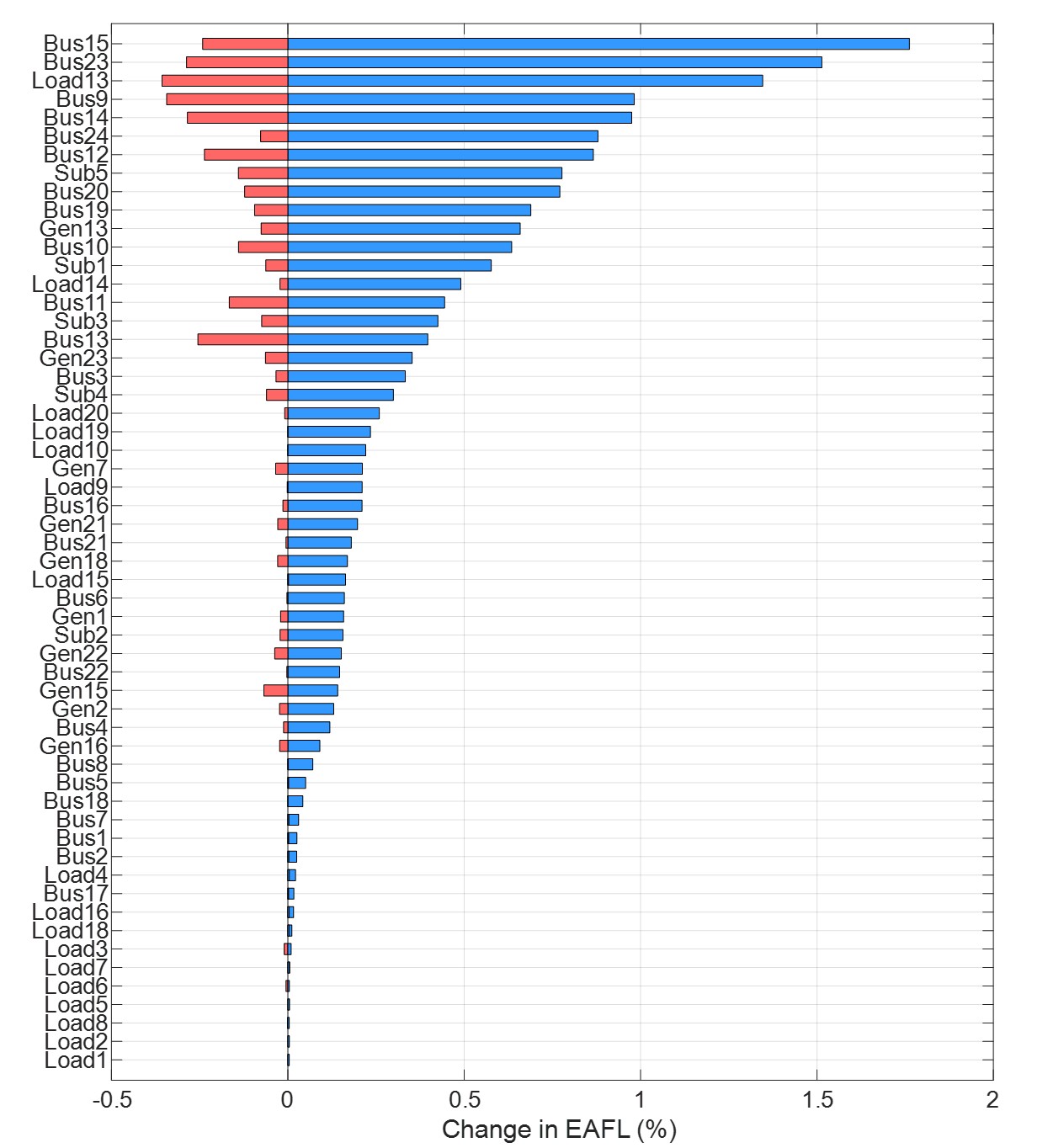}
    \caption{Tornado plot of component-level sensitivity based on OAT fragility perturbations. Components are ranked by absolute change in EAFL (\%).}
    \label{Tornadoplot}
\end{figure}

To prioritize retrofit targets in a computationally efficient manner, a sensitivity analysis is conducted using the OAT approach introduced in Section 2.4. Each EPN component’s median fragility parameter is perturbed individually by a factor of 50\% in turn and the resulting change in the system-level EAFL is recorded, as defined in Equation (\ref{eq18}), representing the marginal impact of the component on system-level risk. The results are presented in Figure \ref{Tornadoplot}, where red bars indicate a reduction in EAFL, i.e., a beneficial effect from the retrofit, while blue bars represent increases in EAFL, e.g., deterioration due to adverse environmental conditions. Components are sorted by the magnitude of their sensitivity index to emphasize those with the greatest influence. As shown, Bus (15, 23, 9, 14, 12, 13), Sub 5, Load 13, and Gen 13 emerge as the most impactful components. These components are thus prioritized in the subsequent optimization phase. It is worth noting that these high-impact components cannot be identified based solely on fragility curves, as their significance also arises from complex interactions among network structure, operational constraints, and hazard intensity uncertainty. This highlights the value of our proposed probabilistic seismic risk assessment framework, which effectively links component vulnerabilities to system-wide outcomes. 

\begin{figure}
    \centering
    \includegraphics[width=0.75\linewidth]
    { 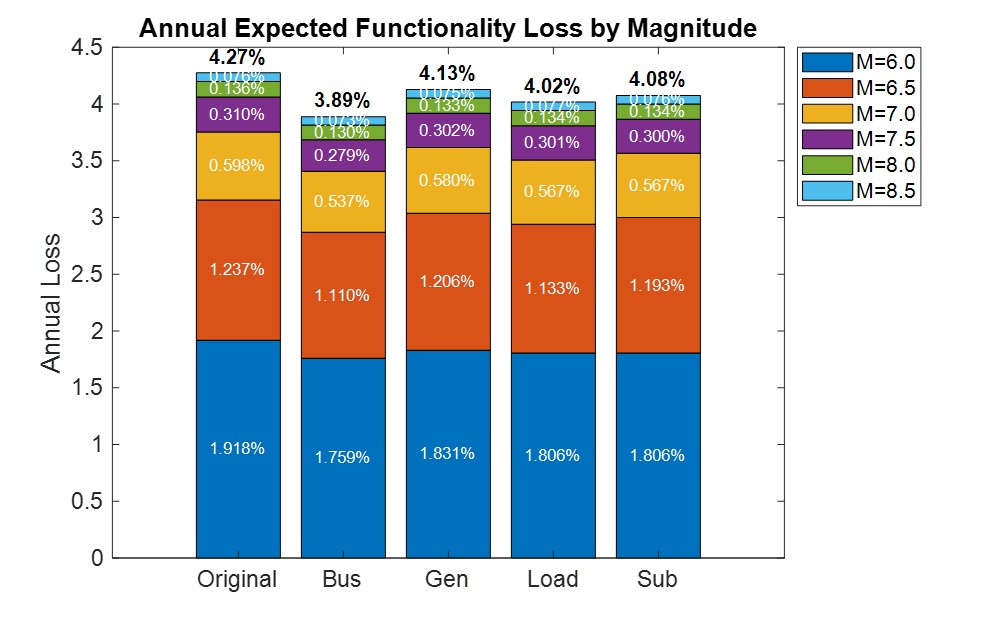}
    \caption{Category-level sensitivity analysis showing the EAFL resulting from upgrading the fragility parameters of all the involved bus nodes, generation plants, load units, and power substations, respectively (see Table \ref{tab:2} for fragility parameter values).}
    \label{Category-level}
\end{figure}

In addition, the results in Figure \ref{Tornadoplot} also reveal that bus nodes exhibit a notably greater influence on EAFL compared to other component types. This is because bus nodes serve as critical hubs in the power network, enabling the transmission of electricity between generation plants, substations, and load units. Their failure often results in topological fragmentation, islanding, and the disconnection of otherwise functional components. In contrast, the failure of a single generator or load unit, while locally impactful, is less likely to disrupt the global system flow unless it coincides with a major bus failure. This observation is further supported by sensitivity analysis at the component category-level. In this analysis, the fragility parameters of all components within a given category—bus nodes, generation plants, load units, or substations—are retrofitted in turn, characterized by upgraded median and standard deviation values from the Hazus Earthquake Model \cite{ulmi2014hazus} (listed in Table \ref{tab:2}). The resulting EAFL is then recalculated to quantify the system-level risk reduction achieved by each category. As shown in Figure \ref{Category-level}, retrofitting bus nodes leads to the most substantial risk reduction, lowering EAFL from the baseline 4.27\% to 3.89\%, compared to 4.13\% for generation plants, 4.02\% for load units, and 4.08\% for substations. Note that upgrading generation plants results in the least reduction in EAFL. This is attributed to two main factors: first, power plants are typically designed with higher seismic performance standards, making them less vulnerable to earthquake-induced damage. Second, power plant failures tend to have localized impacts on power supply capacity but do not directly compromise network topology or trigger large-scale cascading effects. Furthermore, the magnitude-wise decomposition of EAFL reveals that lower-magnitude events, though less destructive individually, dominate the overall annual loss due to their higher occurrence frequency, highlighting the need to consider both hazard intensity and frequency in risk-informed decision-making.

\begin{table}[h]
    \centering
    \caption{Seismic fragility parameters for different EPN components after retrofitting}
    \begin{tabular}{lcccccccc}
        \toprule
        \textbf{Damage} & \multicolumn{2}{c}{\textbf{Bus node}} & \multicolumn{2}{c}{\textbf{Generation plant}} & \multicolumn{2}{c}{\textbf{Load unit}} & \multicolumn{2}{c}{\textbf{Substation}} \\
        \textbf{state} & $\theta(g)$ & $\beta$ & $\theta(g)$ & $\beta$ & $\theta(g)$ & $\beta$ & $\theta(g)$ & $\beta$ \\
        \midrule
        Slight & 0.15 & 0.70 & 0.12 & 0.60 & 0.28 & 0.30 & 0.15 & 0.60 \\
        Moderate & 0.29 & 0.55 & 0.25 & 0.60 & 0.40 & 0.20 & 0.25 & 0.50 \\
        Extensive & 0.45 & 0.45 & 0.52 & 0.55 & 0.72 & 0.15 & 0.35 & 0.40 \\
        Complete & 0.90 & 0.45 & 0.92 & 0.55 & 1.10 & 0.15 & 0.70 & 0.40 \\
        \bottomrule
    \end{tabular}
    \label{tab:2}
\end{table}

Given the practical limitation that not all vulnerable components can be retrofitted due to budget constraints, it is crucial to identify a cost-effective subset whose reinforcement maximizes risk reduction. Building on the prior sensitivity insights, a GA-based optimization is applied to determine the optimal combination of components to retrofit subject to a fixed investment budget. Following the procedure in section 2.4, the GA is configured with the following parameters: a population size of 40, a maximum of 80 generations, and a crossover fraction of 0.8 using the scattered crossover function. Uniform mutation with a mutation rate of 0.1 is applied, and tournament selection is used to retain diversity. Elitism is enabled with 4 top-performing individuals preserved in each generation. The fitness function is defined by a penalized EAFL with an initial penalty factor of 10 (Equation (\ref{eq19})) to ensure solutions remain within the allowable cost range. The optimization is executed in parallel to improve computational efficiency, and convergence progress is tracked using real-time plotting and output logging. To reflect real-world implementation considerations, Table \ref{tab:3} lists the estimated seismic retrofitting costs (in million USD) for each type of EPN component.

\begin{table}[h]
    \centering
    \caption{Seismic retrofitting cost for each EPN component}
    \begin{tabular}{lcccccccc}
        \toprule
        Component & Cost (in million USD) \\
        \midrule
        Bus node & 0.5 \\
        Generation plant & 1.0 \\
        Load unit & 0.3 \\
        Substation & 0.8 \\
        \bottomrule
    \end{tabular}
    \label{tab:3}
\end{table}

\begin{figure}
    \centering
    \includegraphics[width=0.6\linewidth]
    { 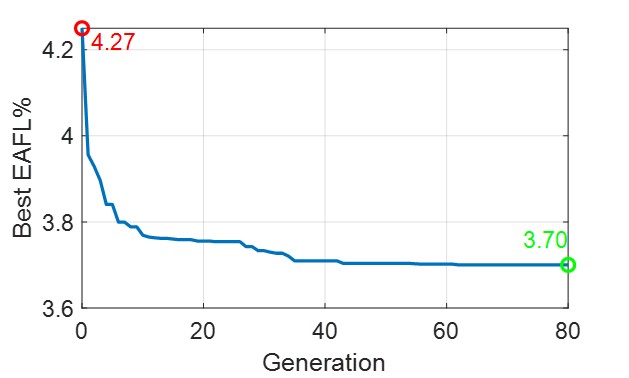}
    \caption{Convergence history of the GA-based retrofit optimization. The blue curve shows the best EAFL (\%) per generation; the red marker (4.27\%) denotes the initial best individual, and the green marker (3.70\%) denotes the final best solution after 80 generations.}
    \label{Convergence_history}
\end{figure}

For demonstration, the available retrofitting budget is first set to 5 million USD, representing a practical but limited investment. Under this constraint, the GA converges to a retrofit configuration consisting of 9 components—six bus nodes (Bus 9, 10, 13, 14, 15, and 23), one load unit (Load 13), and two power substations (Sub 3 and 5)—whose aggregate cost (USD 4.9 million) respects the budget cap. Notably, most of the selected components align with the top-ranked items identified from the sensitivity analysis in Figure 9, reinforcing the reliability of the optimization outcome. Meanwhile, a few retrofitted components (e.g., Sub 3) were not among the most sensitive ones locally, suggesting that the power system exhibits complex interdependencies where the global contribution of a component to overall functionality cannot be fully captured by its individual sensitivity. To this end, our proposed holistic optimization approach effectively captures these complexities and yields cost-efficient retrofit strategies under realistic budget constraints. As demonstrated in Figure \ref{Convergence_history} that displays the evolution trajectory of the best solution: the optimized retrofitting configuration can decrease the EAFL from 4.27\% to 3.70\% by a significant reduction rate of 13.4\%, which outperforms those results in Figure \ref{Category-level} derived from naive or component-centric approaches that may neglect functional interdependencies and spatial hazard exposure.

\subsection{Investment–risk reduction tradeoff and retrofit strategy evolution}
\begin{figure}
    \centering
    \includegraphics[width=0.6\linewidth]
    { 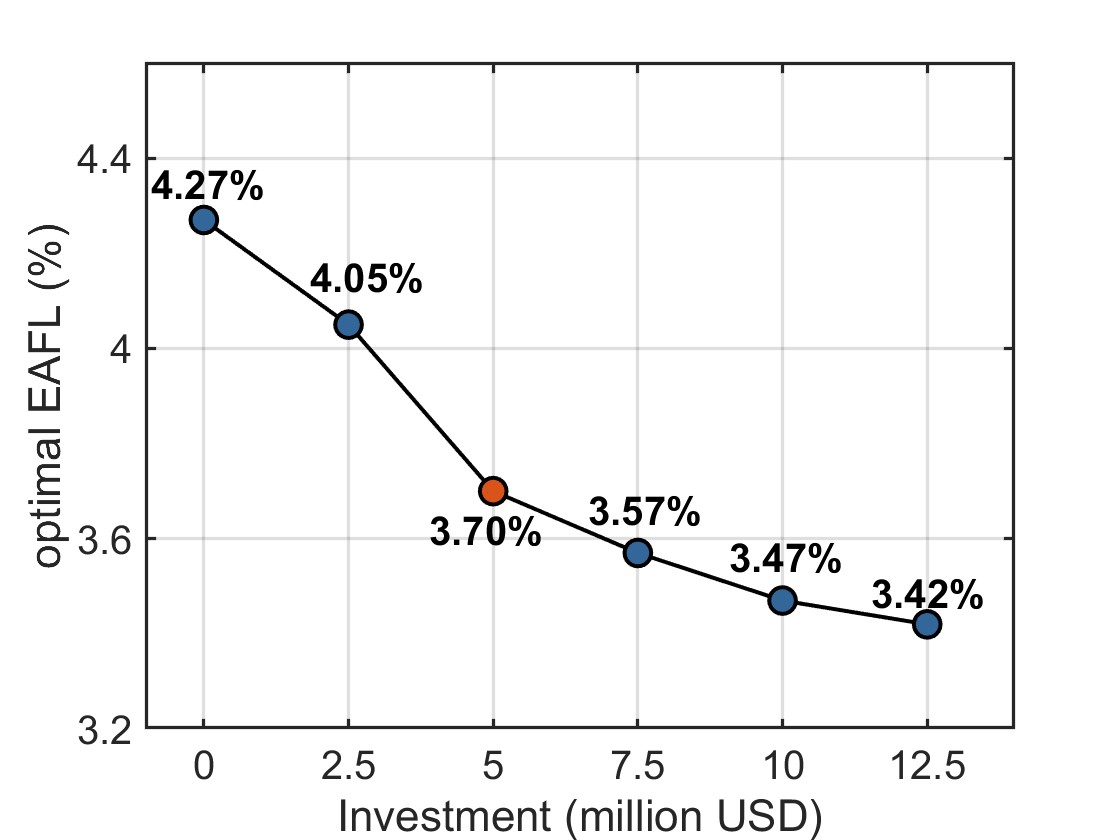}
    \caption{Optimal EAFL under different seismic retrofit investment levels.}
    \label{Optimal_EAFL}
\end{figure}

To explore how seismic risk reduction evolves with increasing investment, a budget–risk sensitivity analysis is performed by solving the retrofit optimization planning problem under varying budget levels ranging from 2.5 to 10.0 million USD. For each specific budget scenario, the GA optimization process is executed independently to identify the optimal retrofit strategy under uncertainty, and the corresponding optimized EAFL is recorded. The results are presented in Figure \ref{Optimal_EAFL}, in which the optimal EAFL-investment curve exhibits a clear diminishing return effect: substantial reductions in EAFL are observed at relatively low investment levels, especially within the first 5.0 million USD marked in orange, whereas additional investments yield progressively smaller improvements. This insight provides practical guidance for decision-makers to balance risk reduction against financial constraints and supports the identification of optimal investment thresholds where the cost-benefit ratio for risk mitigation is most favorable. For instance, here the nonlinear relationship displayed in Figure \ref{Optimal_EAFL} suggests that moderate budgets (around 5.0 million USD) may offer the highest gains in risk reduction per dollar spent.

To illustrate how the optimal retrofit strategies evolve across investment levels, Table \ref{tab:4} summarizes the component-wise retrofit selections for each investment scenario. At the 2.5 million USD investment level, the strategy focuses solely on critical bus nodes (i.e., 9, 13, 14, 15, 23). Increasing the budget to 5.0 million USD expands the strategy to include additional critical bus nodes (10), a high-impact load unit (13), and two vulnerable substations (3, 5). These initial selections at lower budget levels align closely with earlier sensitivity rankings presented in Figure \ref{Tornadoplot}, thereby reinforcing the internal consistency and efficacy of the proposed optimization framework. As the budget increases, the selected components are expected to expand to include lower-ranked but still beneficial candidates. For instance, given 7.5 million USD, the scope of retrofitting expands to include a generation plant (15) and more load units (14, 20), alongside additional bus nodes (12, 20). Notably, at 10.0 million USD, the strategy encompasses a second generation plant (23), more load units (9, 19), and an additional substation (4). This systematic expansion across different component types with increasing budget offers a scalable and incrementally robust roadmap for enhancing the overall performance of the system against earthquakes.

\begin{table}[h]
    \centering
    \caption{Optimized seismic retrofit plans for different investment levels}
    \begin{tabular}{lcccccccc}
        \toprule
        \textbf{Investment} & \multicolumn{4}{c|}{\textbf{Optimal seismic retrofitting configuration}} \\ \cline{2-5} 
        \textbf{(in million USD)} & \textbf{Bus node} & \textbf{Generation plant} & \textbf{Load unit} & \textbf{Substation} \\ 
        \midrule
        2.5 & 9,13,14,15,23 & / & / & / \\ 
        5.0 & 9,10,13,14,15,23 & / & 13 & 3,5 \\ 
        7.5 & 9,10,12,13,14,15,20,23 & 15 & 13,14,20 & 3,5 \\ 
        10.0 & 9,10,12,13,14,15,20,23 & 15,23 & 9,13,14,19,20 & 3,4,5 \\ 
        \bottomrule
    \end{tabular}
    \label{tab:4}
\end{table}

\begin{figure}
    \centering
    \includegraphics[width=1\linewidth]
    { 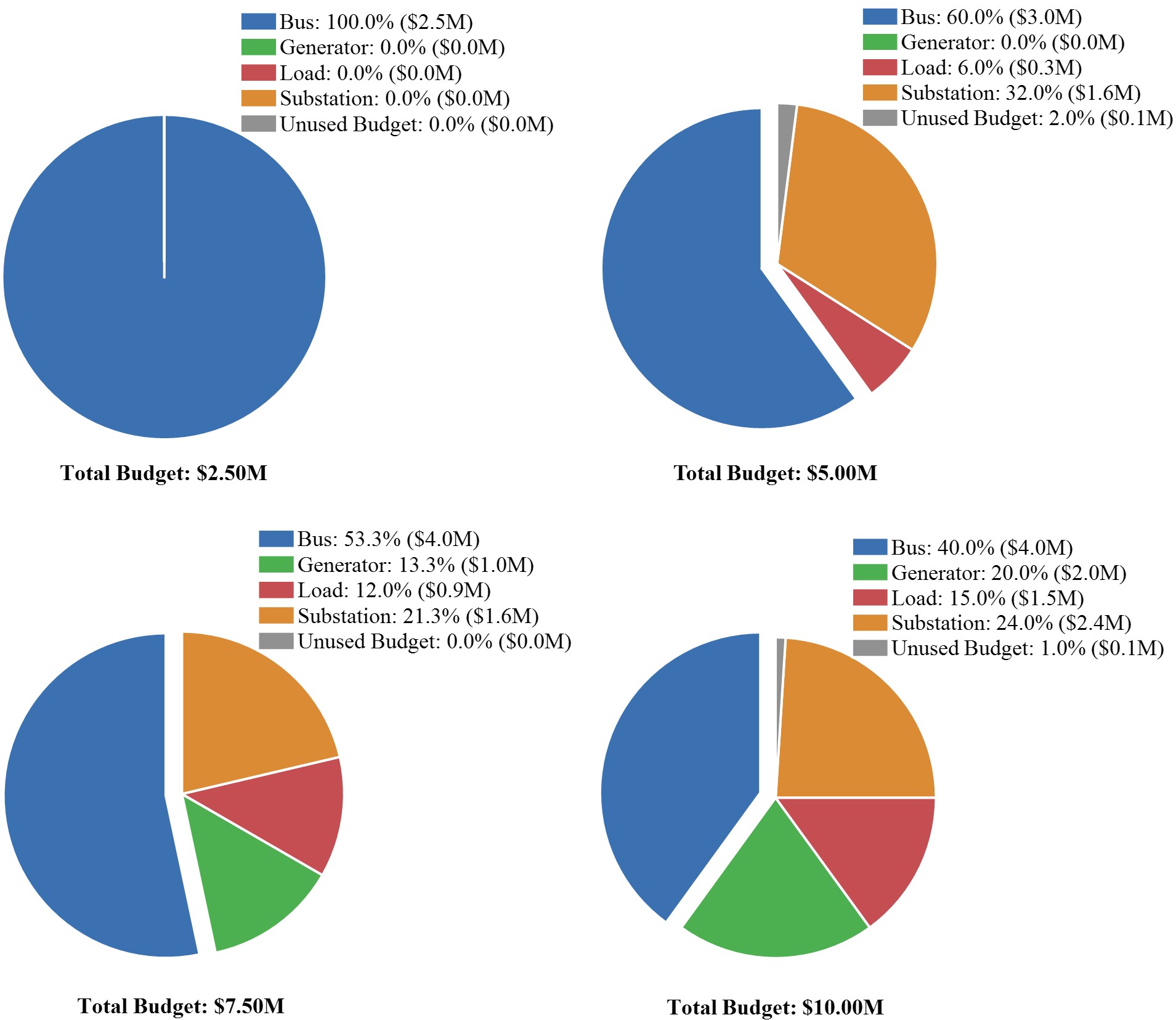}
    \caption{Budget allocation across different component types under various investment levels.}
    \label{Budget_allocation}
\end{figure}

The evolution of budget allocation across component types is further illustrated in Figure \ref{Budget_allocation}, showing how available funds are distributed among bus nodes, generation plants, load units, and substations as the investment level increases. At lower budgets (i.e., 2.5–5.0 million USD), the majority of capitals are allocated to bus nodes, reinforcing their critical role in mitigating cascading failures. As the budget increases, allocations become more diversified, with bus nodes remaining dominant but a significant share going to substations, generation plants, and load units. Together with Figure \ref{Optimal_EAFL}, it illustrates how adaptive and intelligent the proposed framework is in balancing marginal benefits of strengthening various components so that system robustness is progressively enhanced under varied budget constraints. Another notable aspect is the occasional presence of an unused budget (e.g., at 5.0 million USD and 10.0 million USD in Figure \ref{Budget_allocation}). This occurs because the defined budget serves as an upper investment limit, and the optimal combination of retrofit measures identified by the algorithm may not necessarily exhaust the entire available budget. The discrete nature of retrofitting decisions and the specific cost granularity of available upgrade options mean that after prioritizing the most cost-effective interventions, any remaining budget might be insufficient to fund the next beneficial upgrade. This implies that beyond a certain point, the system's inherent design or the currently available retrofitting technologies present limitations to achieving further cost-effective risk reduction, thereby reinforcing the concept of a practical risk floor. Future research could explore alternative retrofit technologies or network reconfigurations to potentially utilize such residual budgets more efficiently.

These findings collectively validate the proposed risk-informed retrofit planning method, demonstrating its ability to produce interpretable, cost-effective retrofitting strategies that prioritize high-impact components while accounting for seismic hazard and component fragility uncertainties, and system topology and operational constraints.

\section{Conclusions}

This study presents a probabilistic framework for seismic risk assessment and retrofit planning of EPNs, addressing critical limitations in existing approaches, including the neglect of system-wide interdependencies, oversimplified binary-state damage assumptions, and omission of electrical operational constraints. The proposed framework integrates spatially correlated seismic hazard modeling, multi-level fragility-based component assessment, graph-based island detection under topological disconnection, DC optimal power flow (DCOPF) analysis under operational constraints, and heuristic optimization under budget limitations. Together, these components enable risk-informed, system-level decision-making under uncertainty. 

The framework is demonstrated on the IEEE 24-bus Reliability Test System, where both scenario-based and probabilistic analyses are performed. The expected annual functionality loss (EAFL) metric is used to quantify seismic risk, and optimal retrofit strategies are identified to minimize this risk within predefined investment limits. The results highlight that moderate investments targeting topologically central assets—especially bus nodes—can lead to significant improvements in system resilience, whereas additional investments yield diminishing returns. This sensitivity-guided optimization offers practical insights into cost-effective retrofit strategies and investment–risk tradeoffs.

Crucially, this framework significantly advances the state-of-the-art by overcoming the fragmented analysis and oversimplified assumptions commonly seen in fragility-only, topology-only, or global DCOPF methods. It achieves higher-fidelity simulations of post-event system functionality evaluation through unified integration. In particular, the coupling of graph-based island detection with DCOPF allows us to effectively capture topological disruptions and dispatch constraints. Without explicitly identifying electrical islands, global DCOPF on disconnected networks often fails to converge due to invalid flows across isolated buses. In the post-earthquake context, isolating and analyzing each island separately ensures feasible and accurate functionality assessments. This advancement provides a more holistic and accurate predictive capability for post-earthquake EPN behavior, a critical gap in existing resilience planning tools. Moreover, the island-based analysis approach provides a valuable perspective for other flow-dependent networked infrastructures facing similar convergence challenges under disruption. 

Owing to its modular and scalable structure, the framework can be extended further to other networks with complex and evolving configurations, customized component fragility data, and site-specific hazard conditions. Future developments may expand this framework to real-world applications and to further incorporate time-dependent recovery processes, multi-hazard scenarios (e.g., earthquake–flood), and co-simulation with interdependent systems such as gas, water, or transportation. Overall, by bridging seismic hazard science, power systems engineering, and decision optimization, this study contributes to advancing resilient infrastructure planning under uncertainty, offering both theoretical insights and practical tools for disaster risk mitigation in critical lifeline systems. 

\section*{Acknowledgments}
The research was conducted at the Singapore-ETH Centre, which was established collaboratively between ETH Zurich and the National Research Foundation Singapore, and CNRS@CREATE through the DESCARTES program, both research supported by the National Research Foundation, Prime Minister’s Office, Singapore under its Campus for Research Excellence and Technological Enterprise (CREATE) programme. B. Moya acknowledges support from the French government, managed by the National Research Agency (ANR), under the CPJ ITTI. Prof. Chinesta acknowledges the support of the Chimera RTE research chairs at Arts et Metiers Institute of Technology (ENSAM). Prof. Chatzi gratefully acknowledges the funding from Horizon Europe under the program HORIZON-CL5-2023-D4-02-01 for project INBLANC - INdustrialisation of Building Lifecycle data Accumulation, Numeracy and Capitalisation, Grant Number: 101147225.

\bibliographystyle{unsrt}  
\bibliography{refs}  

\section*{Appendix}
The predictions for BSSA14 GMPE are given by:
\begin{equation}
    \ln \text{PGA} = F_M + F_D + F_S + \varepsilon
    \label{eq:placeholder}
\end{equation}
The magnitude scaling function FM is expressed as:
\begin{equation}
    F_{x}(M, \text{mech}) = 
    \begin{cases} 
    e_{0}U + e_{1}SS + e_{2}NS + e_{3}RS + e_{4}(M-M_{h}) + e_{5}(M-M_{h})^{2}, & M \leq M_{h} \\
    e_{0}U + e_{1}SS + e_{2}NS + e_{3}RS + e_{6}(M-M_{h}), & M > M_{h} 
    \end{cases}
\end{equation}
The distance attenuation function $F_D$ is given by:
\begin{equation}
    F_D(R, M, region) = [c_1 + c_2(M - M_{\text{ref}})] + (c_3 + \Delta c_3)(R - R_{ref})
\label{eq:my_label}
\end{equation}
where  $R=\sqrt{R_{JB}^2+h^2}$.

The site effect function $F_S$ is expressed as:
\begin{equation}
    F_s(V_{s30}, R_{JB}, M, region, z_1) = \ln(F_{lin}) + \ln(F_{nl})
    \label{eq:example1}
\end{equation}
where
\begin{equation}
    \ln((F_{lin}) = 
    \begin{cases} 
        c \ln(V_{s30} / V_{ref}), & V_{s30} \leq V_{ref} \\
        c \ln(V_c / V_{ref}), & V_{s30} > V_{ref}
    \end{cases}
\label{eq:customLabel}
\end{equation}
\begin{equation}
    \ln(F_{nl}) = f_1 + f_2 \ln \left( \frac{PGA_r + f_3}{f_3} \right)
    \label{eq:placeholder}
\end{equation}
\begin{equation}
    f_2 = f_4 \exp \{ f_5 (\min(V_{s30}, 760) - 360) \} - \exp \{ f_5 (760 - 360) \}
    \label{eq:example}
\end{equation}
The aleatory-uncertainty function is given by:
\begin{equation}
    \sigma(M, R_E, V_{s30}) = \sqrt{\phi^2(M, R_E, V_{s30}) + \tau^2(M)}
    \label{eq:placeholder_label}
\end{equation}
where
\begin{equation}
    \phi(M, R_{JB}, V_{s30}) = 
    \begin{cases} 
        \phi(M, R_{JB}), & V_{s30} > V_{2} \\
        \phi(M, R_{JB}) - \Delta \phi_{v} \frac{\ln(V_{2} / V_{s30})}{\ln(V_{2} / V_{1})}, & V_{1} \leq V_{s30} \leq V_{2} \\
        \phi(M, R_{JB}) - \Delta \phi_{v}, & V_{s30} \leq V_{1} 
    \end{cases}
    \label{eq:phi_vs30}
\end{equation}
\begin{equation}
\phi(M, R_{JB}) = 
\begin{cases} 
\phi(M) & \quad R_{JB} \leq R_1 \\
\phi(M) - \Delta \phi_R \frac{\ln(R_{JB} / R_1)}{\ln(R_2 / R_1)} & \quad R_1 < R_{JB} \leq R_2 \\
\phi(M) - \Delta \phi_R & \quad R_{JB} > R_2 
\end{cases}
\label{eq:phi_M_R_JB}
\end{equation}
\begin{equation}
\begin{cases} 
\phi(M) = 0.495 \\
\tau(M) = 0.348 \quad \text{if } M \geq 5.5 
\end{cases}
\label{eq:placeholder}
\end{equation}
The associated model coefficients corresponding to a Period of 0 second are determined by referring to \cite{boore2014nga} and summarized as below.

\begin{table}[h]
    \centering
    \caption{Placeholder caption for model coefficients of the BSSA14 GMPE.}
    \begin{tabular}{cccccccc}
        \hline
        \multicolumn{8}{c}{Magnitude Scaling} \\
        \hline
        $e_0$ & $e_1$ & $e_2$ & $e_3$ & $e_4$ & $e_5$ & $e_6$ & $M_h$ \\
        \hline
        0.4473 & 0.4856 & 0.2459 & 0.4539 & 1.431 & 0.05053 & -0.1662 & 5.5 \\
        \hline
    \end{tabular}

    \begin{tabular}{cccccc}
        \hline
        \multicolumn{6}{c}{Distance Scaling} \\
        \hline
        $c_1$ & $c_2$ & $c_3$ & $M_{\text{ref}}$ & $R_{\text{ref}}$ & $h$ \\
        \hline
        -1.134 & 0.1917 & -0.00809 & 4.5 & 1 & 4.5 \\
        \hline
    \end{tabular}

    \begin{tabular}{cccccccc}
        \hline
        \multicolumn{8}{c}{Site Term} \\
        \hline
        $\Delta c_3$ & $c$ & $V_c$ & $V_{\text{ref}}$ & $f_1$ & $f_3$ & $f_4$ & $f_5$ \\
        \hline
        0.00286 & -0.5150 & 925.0 & 760 & 0 & 0.1 & -0.1500 & -0.00701 \\
        \hline
    \end{tabular}

    \begin{tabular}{cccccc}
        \hline
        \multicolumn{6}{c}{Aleatory Uncertainty} \\
        \hline
        $R_1$ & $R_2$ & $\Delta \phi_R$ & $\Delta \phi_V$ & $V_1$ & $V_2$ \\
        \hline
        110.0 & 270.0 & 0.100 & 0.084 & 225 & 300 \\
        \hline
    \end{tabular}
    \label{tab:placeholder_label}
\end{table}

\end{document}